\documentclass[xcolor=dvipsnames]{aa}  
\usepackage{placeins}

\usepackage{amsmath}
\usepackage{graphicx}
\usepackage{adjustbox}
\usepackage{txfonts}
\def\be{\begin{equation}}
\def\ee{\end{equation}}

\usepackage{xcolor}

\newcommand{\srg}{\textit{SRG}}

\DeclareUnicodeCharacter{2212}{\textendash}

\defcitealias{2021A&A...651A..41C}{Paper~I}

\begin{document}

   \title{Tempestuous life beyond $R_{500}$: \\X-ray view on the Coma cluster with SRG/eROSITA}

   \subtitle{II. Shock \& relic}
   
   \author{E. Churazov \inst{1,2} 
   \and
   I. Khabibullin \inst{3,2,1}
   \and
   A.M.Bykov \inst{4}
   \and
   N.Lyskova \inst{2,5}
   \and
   R.Sunyaev \inst{2,1}}
   
\institute{Max Planck Institute for Astrophysics, Karl-Schwarzschild-Str. 1, D-85741 Garching, Germany 
\and 
Space Research Institute (IKI), Profsoyuznaya 84/32, Moscow 117997, Russia
\and
Universitäts-Sternwarte, Fakultät für Physik, Ludwig-Maximilians-Universität München, Scheinerstr.1, 81679 München, Germany
\and
Ioffe Institute, Politekhnicheskaya st. 26, Saint Petersburg 194021, Russia
\and
ASC of P.N.Lebedev Physical Institute, Leninskiy prospect 53, Moscow 119991, Russia
}

   \date{Received ; accepted }

  \abstract
{This is the second paper in a series of studies of the Coma cluster using the \textit{SRG}/eROSITA X-ray data obtained during the calibration and performance verification phase of the mission. Here, we focus on the region adjacent to the radio source 1253+275 (radio relic, RR, hereafter). We show that the X-ray surface brightness exhibits its steepest gradient at $\sim 79'$ ($\sim 2.2\,{\rm Mpc}\approx R_{200c}$), which is almost co-spatial to the outer edge of the RR. As in the case of several other relics, the Mach number of the shock derived from the X-ray surface brightness profile ($M_X\approx 1.9$) appears to be lower than needed to explain the slope of the integrated radio spectrum in the diffusive shock acceleration (DSA) model ($M_R\approx 3.5$) if the magnetic field is uniform and the radiative losses are fast. However, the shock geometry is plausibly much more complicated than a spherical wedge centered on the cluster, given the non-trivial correlation between radio, X-ray, and SZ images.  While the complicated shock geometry alone might cause a negative bias in $M_X$, we speculate on a few other possibilities that may affect the $M_X$-$M_R$ relation, including the shock substructure that might be modified by the presence of non-thermal filaments stretching across the shock and the propagation of relativistic electrons along the non-thermal filaments with a strong magnetic field. We also  discuss the "history" of the radio galaxy NGC4789, which is located ahead of the relic in the context of the Coma-NGC4839 merger  scenario.
}


   \maketitle
%



\section{Introduction}
The Coma cluster (Abell 1656) is one of the nearby ($z=0.0231$) massive ($M_{500}\sim 6\,10^{14}\,M_\odot$; \citealt{2013A&A...554A.140P}) clusters currently experiencing a merger with the NGC~4839 group and which might be close to the apocenter of its orbit \citep[e.g.,][]{1994ApJ...427L..87B,2019MNRAS.485.2922L,2019ApJ...874..112S}. Coma is particularly well known because many phenomena associated with galaxy clusters have been first identified in this cluster. For instance, Coma was proposed as the first target for detection of the SZ effect \citep{1970Ap&SS...7....3S} and was used as a target during the first attempt to detect the SZ shadow \citep{1973SvA....16.1048P}. In addition, an extended radio source 1253+275 was discovered in Coma \citep[e.g.,][]{1979ApJ...233..453J,1985A&A...150..302G}. This source became one of the founding members of the radio relic (hereafter, RR) class of sources. These objects, characterized by their steep spectra (integrated spectral index $\alpha \gtrsim 1$), are often found in the outskirts of clusters and believed to be associated with shocks \citep[see][for a recent review]{2019SSRv..215...16V}. The RR in Coma lies some 70-$80'$ from the center in projection, roughly near the radius $R_{200c}\approx 2\,{\rm Mpc}$, where the enclosed mean total matter density is a factor of 200 higher than the critical density of the Universe.  The shocks that give rise to relics should be visible in X-ray or SZ images, providing complementary constraints on the properties of the shock \citep[see][for the analysis of \textit{XMM-Newton}, \textit{Suzaku}, and \textit{Planck} data on the Coma cluster]{2006A&A...450L..21F,2013MNRAS.433.1701O,2013PASJ...65...89A,2013ApJ...775....4S,2015MNRAS.447.2497E,2016AA...591A.142B}. However, robust characterizations of the X-ray emission at $R_{200c}$ is difficult due to the faintness of the cluster emission relative to the detector and astrophysical backgrounds and foregrounds. Here,  we use the data of the eROSITA telescope on board the \srg~ observatory obtained during the performance verification and calibration phase of the mission.        

Throughout the paper, we assume a flat Lambda Cold Dark Matter ($\Lambda$CDM) cosmology with $\Omega_m = 0.3$, $\Omega_{\Lambda}$=0.7, $H_0 = 70$ km/s/Mpc. At the redshift of the Coma cluster, $1'$ corresponds to 27.98 kpc. We assume $r_{500c}\approx 47'$ \citep{2013A&A...554A.140P} and $r_{200c} \simeq 1.5 r_{500c} \simeq 70'$ for the concentration parameter $c\simeq4$ typical for massive galaxy clusters \citep[e.g.,][among others]{2016MNRAS.457.4340K}. For all radial profiles, we fix the Coma center  at (RA, Dec.) = (12$^h$59$^m$47$^s$, +27$^{\circ}$55'53"), as done in \cite{2013A&A...554A.140P}.

\section{Brief summary of previous studies of the relic region in X-rays and SZ}
\cite{2006A&A...450L..21F} analyzed XMM-Newton observation of the Coma radio relic and suggested that turbulence rather than a shock might be  responsible for the origin of radio emission at the relic position. \cite{2013MNRAS.433.1701O} re-analyzed the XMM-Newton data, using an additional observation in the SW direction as background, and found a temperature jump at the outer edge of the relic, suggesting the presence of a shock wave with the Mach number of $1.9^{+0.16}_{-0.40}$.

\cite{2013ApJ...775....4S} analyzed a large mosaic of Suzaku observations of the Coma Cluster. The measured surface brightness profile along the SW direction drops dramatically approximately at the relic position. \cite{2013ApJ...775....4S} detected no cluster emission  beyond the relic, so only a lower limit  on  the  surface  brightness  jump  (a  factor  of  $\sim$ 13) was obtained.  \cite{2013PASJ...65...89A} analyzed the same data but   used a different observation to model the CXB. These latter authors discovered steep drops both in temperature and X-ray surface brightness across the relic, which are consistent with the $M = 2.2 \pm 0.5$ shock wave.

Based on the analyses of the first public Planck  all-sky data release, \cite{2015MNRAS.447.2497E} constructed a pressure profile in the direction of the radio relic and identified a pressure discontinuity at the relic position, corresponding to the  $M=2.9^{+0.8}_{-0.6}$ shock. Later, using the 2015 Planck data release and assuming different shock geometry,  \cite{2016AA...591A.142B} confirmed the presence of a shock with $M = 2.2 \pm 0.3$.

\section{Observations and data processing}
The \srg~ X-ray observatory \citep{2021A&A...656A.132S}  was launched on July 13, 2019  from the Baikonur cosmodrome. It carries two wide-angle grazing-incidence X-ray telescopes, namely, eROSITA \citep{2021A&A...647A...1P} and the Mikhail Pavlinsky ART-XC telescope \citep{2021A&A...650A..42P}, which operate in the overlapping energy bands of 0.2–8 and 4–30 keV, respectively.

The main dataset used here is similar to the one described in \cite{2021A&A...651A..41C}, hereafter noted as Paper I. In brief, dedicated \textit{SRG} observations of the Coma cluster were performed in two parts, taking place on December 4-6, 2019 and June 16-17, 2020 in the frame of the Calibration and Performance Verification phase (hereafter, CPV). The observatory was operating in a scanning mode, when a rectangular region of the sky is scanned multiple times to ensure the uniform exposure of the region. In addition, we used the data from the all-sky survey in order to determine the typical sky background level around the cluster.

\section{Images and removal of stray light and background sources}
\label{sec:images}
\begin{figure}
\centering
\includegraphics[angle=0,trim= 0mm 0cm 0mm 0cm,width=0.99\columnwidth]{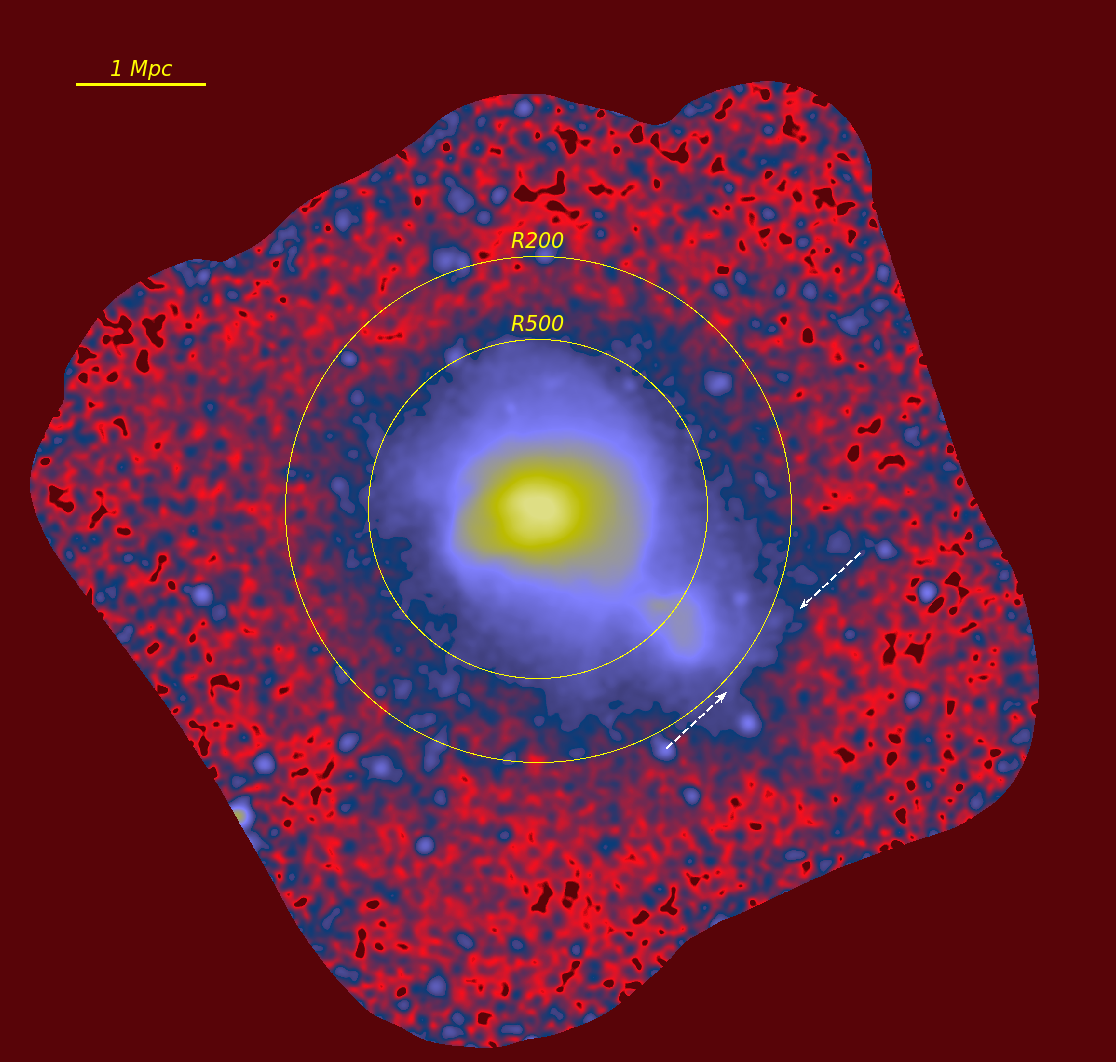}
\caption{Coma cluster $5\times 5$ deg image in the 0.4-2.3 keV band based on CPV data. Individual bright sources have been removed from the image. The resulting image (detector background subtracted and vignetting corrected) is   smoothed with a $\sigma=60''$ Gaussian. The two circles show the estimated positions of $R_{500c}$ and $R_{200c}$. The excess emission at $r\sim R_{500c}$ in the SW direction is very clear. Indeed, in the majority of other directions, a similar level of X-rays  surface brightness is more characteristic for $r\approx R_{500c}$. 
Two arrows show the position of the 1253+275 radio source (its brightest part). A surface brightness edge is visible close to the position of the relic. 
}
\label{fig:coma_s60}
\end{figure}
The Coma cluster X-ray 0.4-2.3~keV image smoothed with a $\sigma=60"$ Gaussian is shown in Fig.~\ref{fig:coma_s60}. Bright compact sources were removed before smoothing. The removal of bright sources was done as in Paper I. 

At large distances from the cluster center (up to $\sim3$ degrees), the surface brightness is very low and might be affected by the telescope's stray light. To subtract the contribution of the stray light, a special procedure was developed,  described in Appendix~\ref{app:stray}. The image shown in   Fig.~\ref{fig:coma_s60} has already been corrected for the stray light contribution.

The two circles in this image show the estimated positions of $R_{500c}$ and $R_{200c}$. A clear excess in the direction of  NGC4839 (to the SW of the Coma core) is present. In this area, the surface brightness near $R_{200c}$ is similar to the surface brightness at  $\sim R_{500c}$ in other directions.  Beyond $R_{200c}$, the surface brightness declines sharply.

\begin{figure*}
\centering
\includegraphics[angle=0,trim= 0mm 0cm 0mm 0cm,width=0.99\columnwidth]{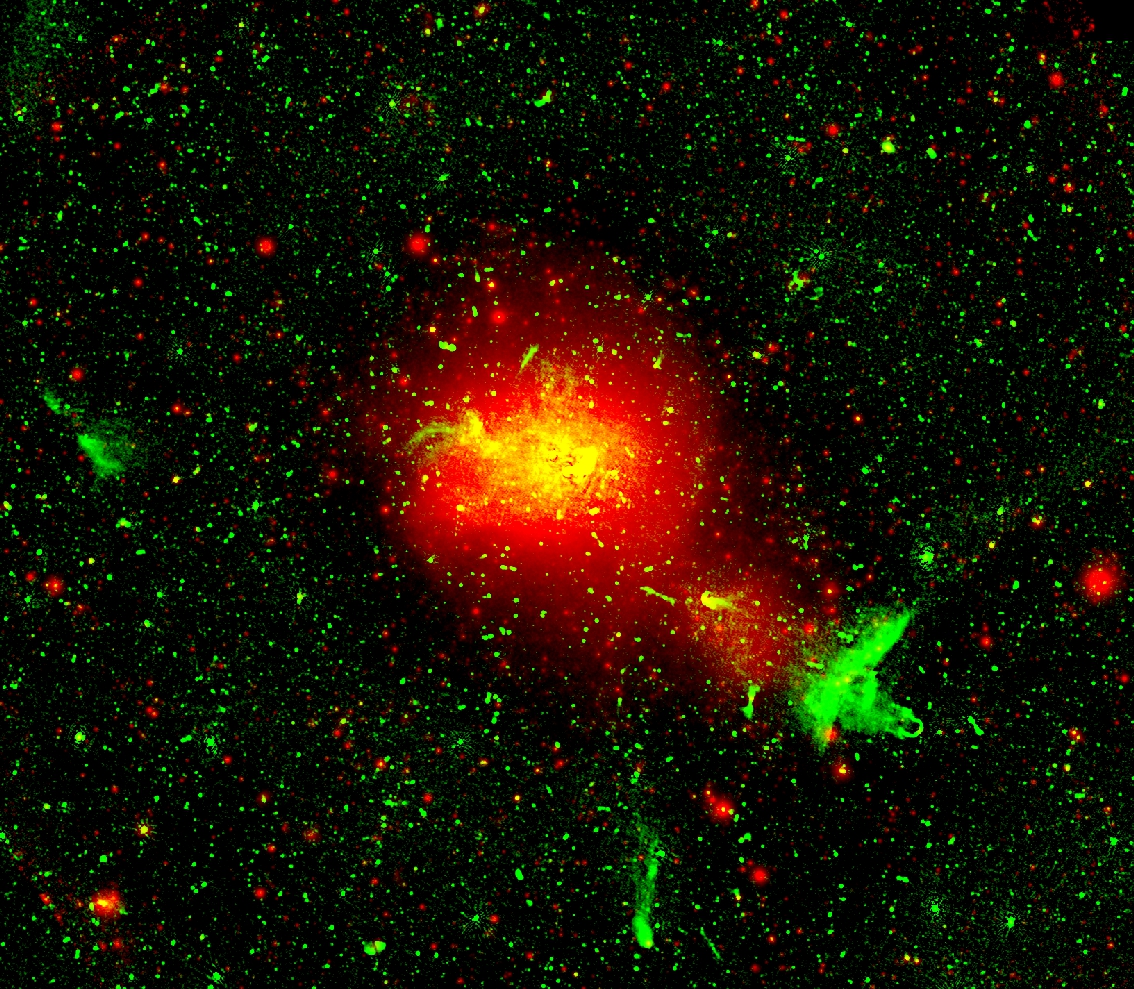}
\includegraphics[angle=0,trim= 0mm 0cm 0mm 0cm,width=0.99\columnwidth]{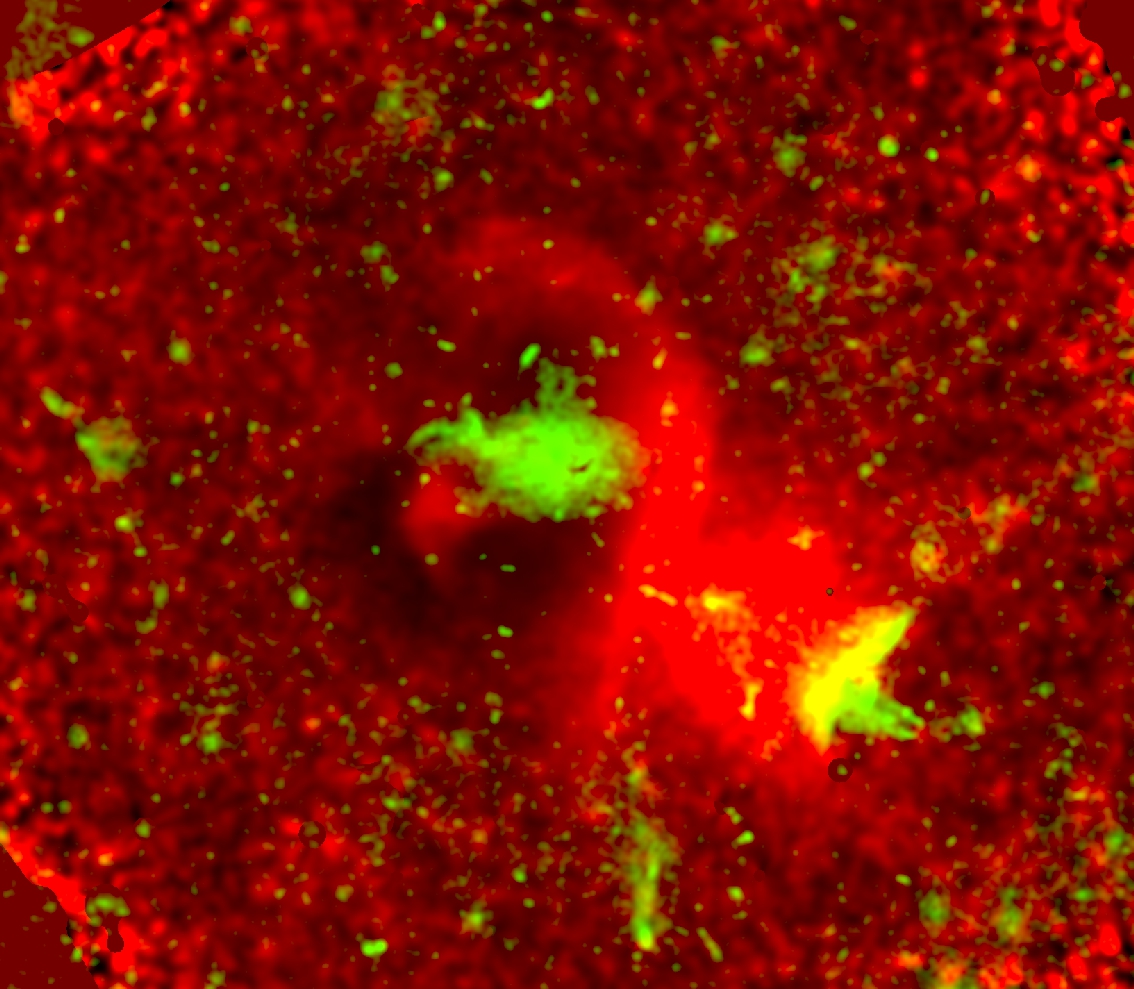}
\caption{Composite X-ray and radio images: eROSITA (red) + LOFAR (green), shown on the left. The LOFAR data are from the publicly available DR2 set \citep{2022A&A...659A...1S}, which is missing the shortest baselines so that extended structures like the Coma radio-halo are partly filtered out. For this reason, the yellowish color corresponding to X-ray and radio bright regions is confined to the bright structures in the core. SW of the core, the red blob is the X-ray emission of the hot gas in the NGC4839 group. The yellow structure inside the group is the radio emission from NGC4839 itself, powered by its central AGN. Further to the SW are the radio relic and the NGC4789 radio galaxy.
{\bf } Details of the right panel are the same as in the left, except that eROSITA image is now divided by the sum of the best-fitting $\beta-$model plus the sky background to emphasize the asymmetric features. The LOFAR image was smoothed after masking bright compact sources. In this representation, the relic edge is aligned with the weak diffuse X-ray structure that extends into the NW direction beyond the brightest part of the edge of the radio relic. 
}
\label{fig:coma_erosita_lofar}
\end{figure*}

\begin{figure*}
\centering
\includegraphics[angle=0,trim= 0mm 0cm 0mm 0cm,width=1\columnwidth]{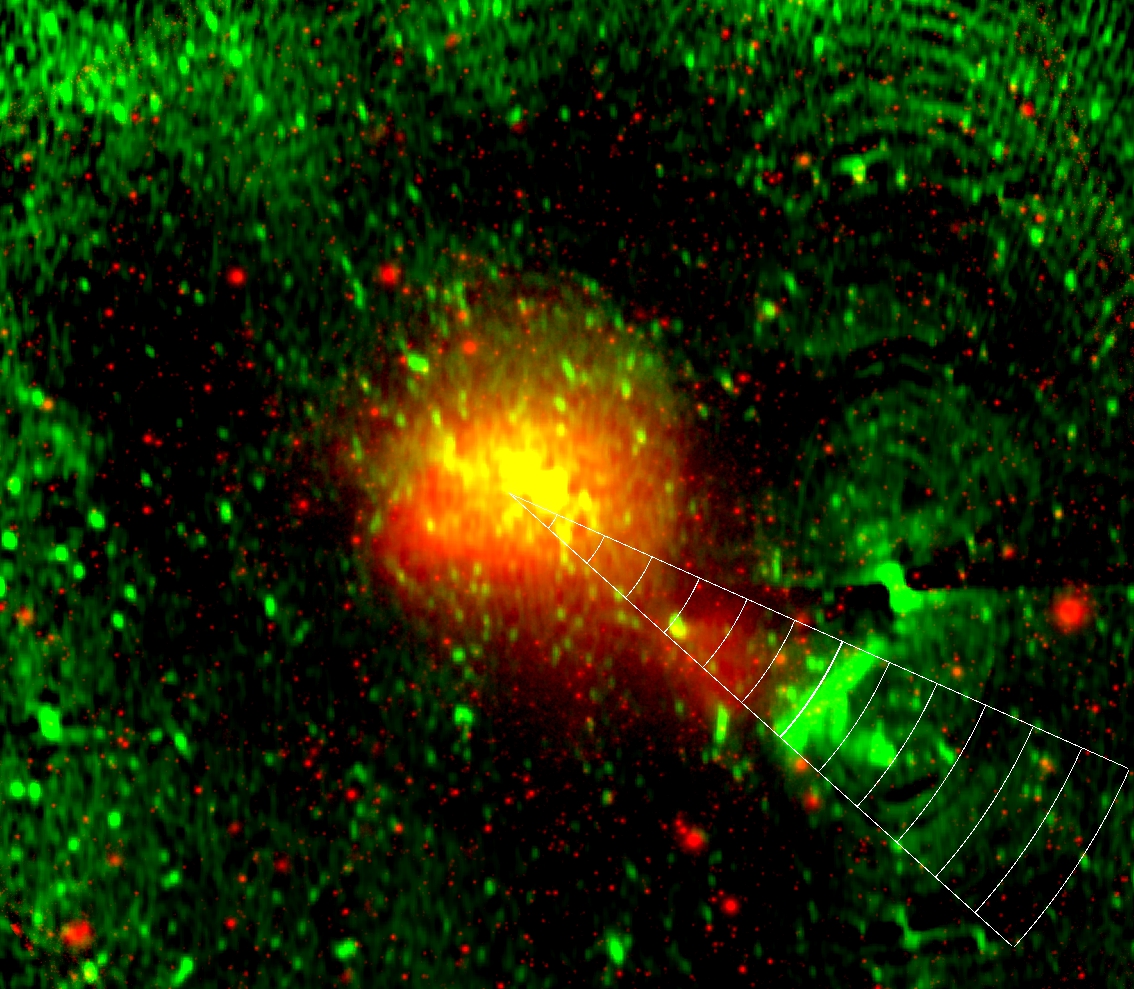}
\includegraphics[angle=0,trim= 0mm 0cm 0mm 0cm,width=1\columnwidth]{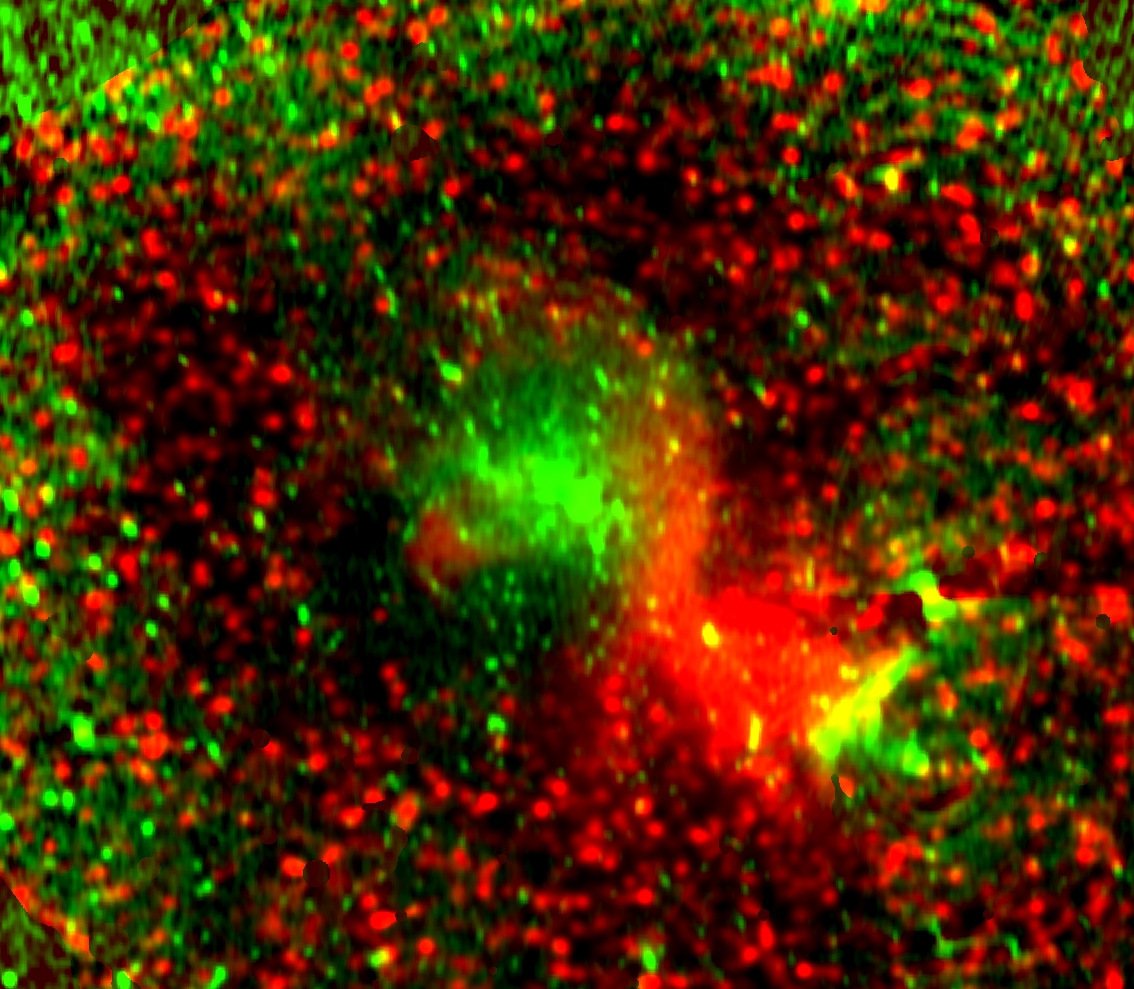}
\caption{Composite X-ray and radio images, similar to Fig.~\ref{fig:coma_erosita_lofar}, except for radio data, which are from the WSRT observations at 325~MHz  \citep{2011MNRAS.412....2B}. In the {\bf left panel,} original images are used. The yellow color in the center reflects the co-spatial X-ray and radio cores of Coma. The red region to the SW of the center corresponds to the dense gas of the NGC4839 group. The radio source NGC4839 is seen as a bright green or yellow patch that is $\sim 40'$ from the center. Further to the SW, a bright green region at $70-80'$ from the core is the radio relic. A wedge containing the relic is shown in white color with $10'$ steps in radius.
In the {\bf right panel}, the X-ray image is divided by the best-fitting beta model (plus sky background) to emphasize the faint structure beyond the NGC4839 group and near the relic. In this image, a red spot that is $20-30'$ to the SE of the core corresponds to the radio quiet area associated with a contact discontinuity. To the west of the core, a shock at $\sim 30'$ is seen as a long arc. To the SW from the core, the X-ray bright region extends all the way to the radio relic, which has a mostly yellow color, indicating co-spatial radio and (faint) X-ray emission distribution. A green feature located further to the SW is a tailed radio galaxy NGC4789.}
\label{fig:xr}
\end{figure*}

To better illustrate the correspondence between structures visible in X-ray and radio bands, we show two-color combinations of X-ray  (in red) and radio (in green) images  in  Figs.~\ref{fig:coma_erosita_lofar} and \ref{fig:xr}. In these figures, we used the LOFAR 144~MHz images from DR2 \citep{2022A&A...659A...1S}, the WSRT image at 352~MHz from \cite{2011MNRAS.412....2B}, and the eROSITA 0.4-2.3~keV image after removing bright point sources and light smoothing (left panels in Figs.~\ref{fig:coma_erosita_lofar} and \ref{fig:xr}). In the right panels of Figs.~\ref{fig:coma_erosita_lofar} and \ref{fig:xr}, the flattened  eROSITA images are used. Namely, the eROSITA image was divided by the sum of the best-fitting $\beta-$model plus the sky background to (i) emphasize asymmetric features and  deviations from this simple radial profile, while (ii) avoiding too much noise amplification at the image edges.

The left panels of Figs.~\ref{fig:coma_erosita_lofar} and  \ref{fig:xr}  clearly show the three main structures: (i) the cluster core, which is bright in X-ray and radio bands, (ii) the X-ray emission of the NGC4839 group, which hosts a radio bright core, and (iii) the relic itself, some 70-80$'$ from the Coma center. We note  that publicly available LOFAR DR2 images lack the largest scales (shortest baselines) \citep{2022A&A...659A...1S} and the extended Coma radio-halo is partly filtered out \citep[see][for the less filtered version of the halo image]{2021ApJ...907...32B,2022ApJ...933..218B}. 

In the right panels of Figs.~\ref{fig:coma_erosita_lofar} and \ref{fig:xr}, several other X-ray features become clearly visible: a) a "red" X-ray-bright blob without prominent radio emission to the SE off the core, b) a bright arc some $30'$ to the West off the core, associated with the shock, and c) the gas of the NGC4839 group to the SW off the core, which joins the relic region further to the SW. The bulk of the relic appears yellow in these images, meaning that both radio and X-ray emissions are present.

A more subtle, but intriguing feature is a fainter X-ray emission to the NW from the brightest part of the relic (see Fig.~\ref{fig:coma_s60}) with a similarly faint radio emission, which is hardly visible in the WSRT image shown in Fig.~\ref{fig:xr}. This feature has been discussed by \cite{2011MNRAS.412....2B} as a part of the "extended relic" and might be visible in the LOFAR images at 144~MHz \citep{2021ApJ...907...32B}, although it is not clear what part of this structure is robust \citep{2022ApJ...933..218B}.  There is also an excess of the tSZ signal approximately in the same region \citepalias[see fig.8 in][]{2021A&A...651A..41C}. We return to this issue in the discussion section.  

\section{Radial profiles}
\label{sec:radial}
\begin{figure}
\centering
\includegraphics[angle=0,trim= 0mm 5cm 0mm 3cm,width=1\columnwidth]{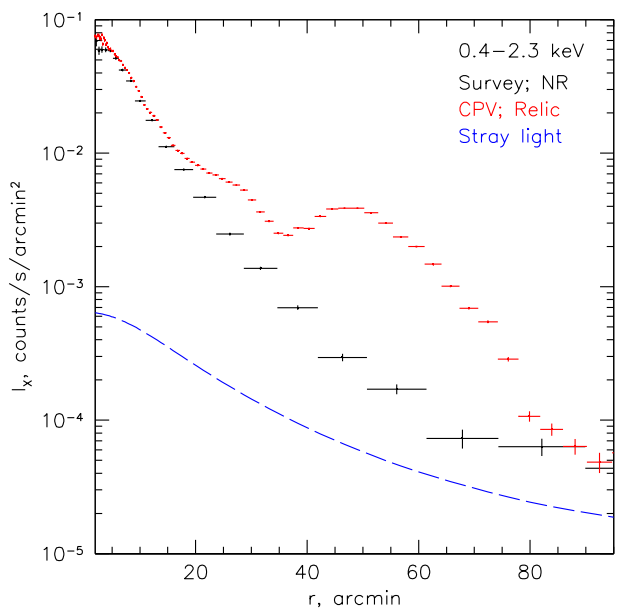}
\caption{X-ray surface brightness profile (0.4-2.3~keV) in the relic (red) and "non-relic" (black) wedges. The detector and sky backgrounds, including stray light, which is shown by the blue dashed line, have been subtracted.
Here, the relic wedge corresponds to 318-336 degrees (counted from the west), while the non-relic wedge covers 0-270 degrees. For the non-relic wedge, the data from the all-sky survey were used. A prominent "bump" at 40-60$'$ in the relic wedge is due to the gas of the NGC4839 group. A steep gradient at $\sim 80'$ is seen in the relic wedge (red crosses). Beyond this radius, the surface brightness in the relic wedge matches approximately the mean surface brightness of Coma in other directions.
 }
\label{fig:radial_relic}
\end{figure}

The surface brightness profiles in the 0.4-2.3~keV were extracted in two wedges. One wedge is in the direction of the relic (318-336 degrees, counted from the west; see the left panel in Fig.~\ref{fig:xr}). For reference, we extracted a non-relic ("NR") radial profile in another wedge (0-270 degrees), which characterizes the mean Coma radial profile in other directions, not affected by the NGC4839 group or the relic itself. These profiles are shown in Fig.~\ref{fig:radial_relic}. The contributions of the detector and sky backgrounds (including stray light) have been subtracted. Its estimated contribution is shown with the blue dashed line.

While the non-relic profile is smooth and monotonic, three distinct features are clearly seen in the relic wedge. Between $\sim 20'-35'$, there is a flattening of the surface brightness followed by its steep decrease. These structures are clearly associated with the shock enveloping the Coma core from the West. The next one is a prominent "bump" at $\sim 50'$ that is associated with the gas of the NGC4839 group. The "bump" continues up to $\sim 70'$, followed by a sharp decrease of the surface brightness between $70$ and $80$ arcminutes, where the relic is located.  Beyond $80'$, the surface brightness in the relic and non-relic wedges are comparable.


\begin{figure}
\centering
\includegraphics[angle=0,trim= 0mm 5cm 0mm 3cm,width=1\columnwidth]{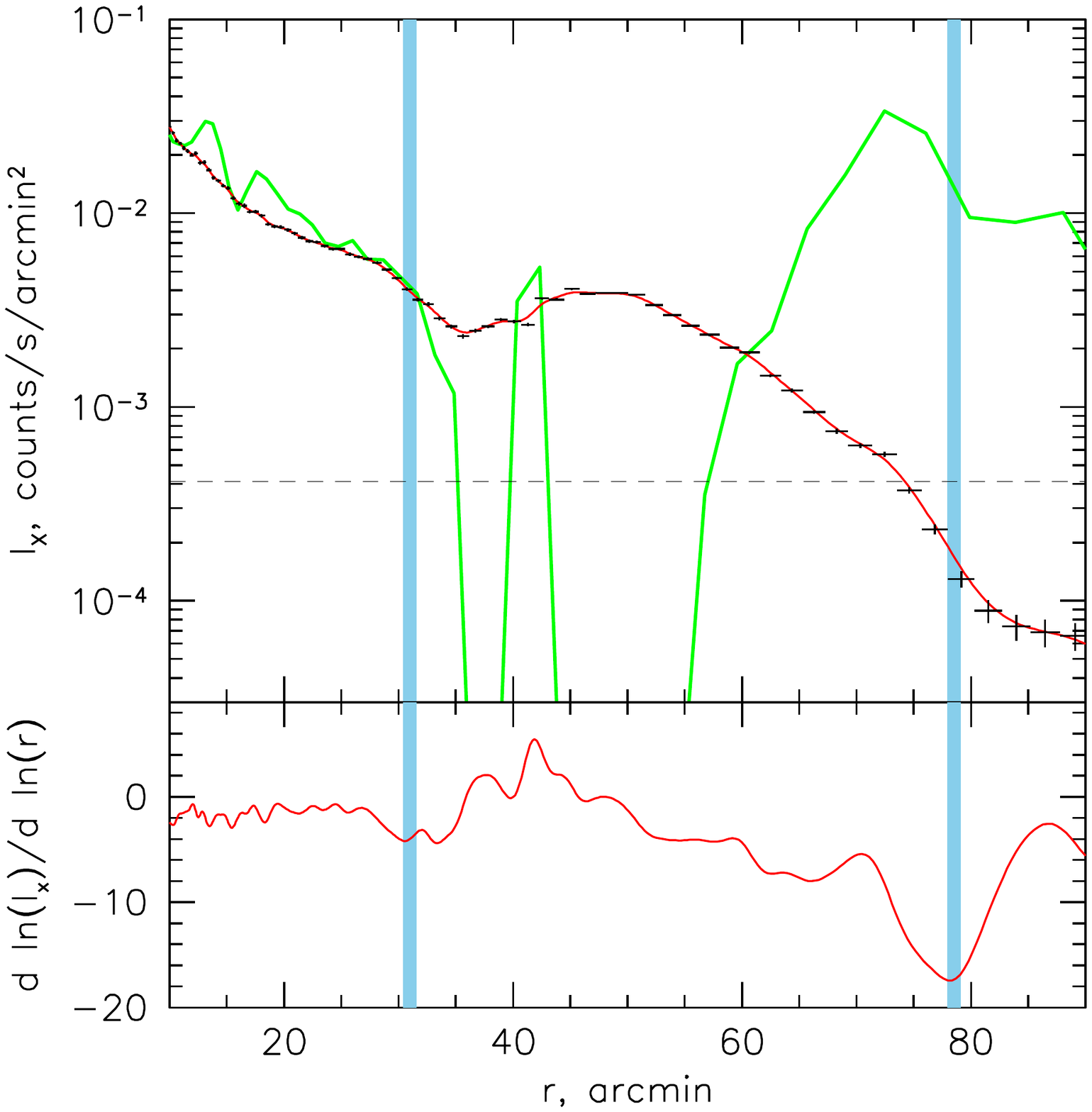}
\caption{Comparison of X-ray and radio radial profiles in the relic wedge. {\bf Top panel} shows the  X-ray surface brightness profile in the 0.4-2.3~keV band (black points). The red line shows the interpolated profile. For comparison, the green line shows the WSRT profile in the same wedge extracted from the 325~MHz image of \cite{2011MNRAS.412....2B}. The radio profile was renormalized to match the X-ray profile in the Coma core. The {\bf bottom panel} shows the logarithmic derivative of the interpolated X-ray surface brightness profile. Two regions with the steepest negative gradients are marked with blue vertical lines. The inner line corresponds to the $30'$-shock, while the outer one - to the relic. }
\label{fig:slope_relic}
\end{figure}

To further illustrate the correspondence between the X-ray and radio structures, Fig.~\ref{fig:slope_relic} compares radial profiles in the relic wedge in both bands. The radio data (WSRT; the green line) have been scaled to match approximately the surface brightness in the Coma core. The radio emission broadly follows the X-ray emission in the core up to $\sim30'$. At larger radii, there is an isolated "bump" at $r\sim 40'$, which is due to radio galaxy NGC4839. The next prominent bump at $70'-80'$ corresponds to the relic region.

A logarithmic derivative of the X-ray surface brightness profile is shown in the bottom panel of Fig.~\ref{fig:slope_relic}. There are two regions with the most prominent negative gradients - one at $30'$ (Western shock) and another one near $80'$ (Relic region), which are marked with two vertical blue lines.

\begin{figure}
\includegraphics[trim=0cm 5cm 0cm 3cm,width=1\columnwidth]{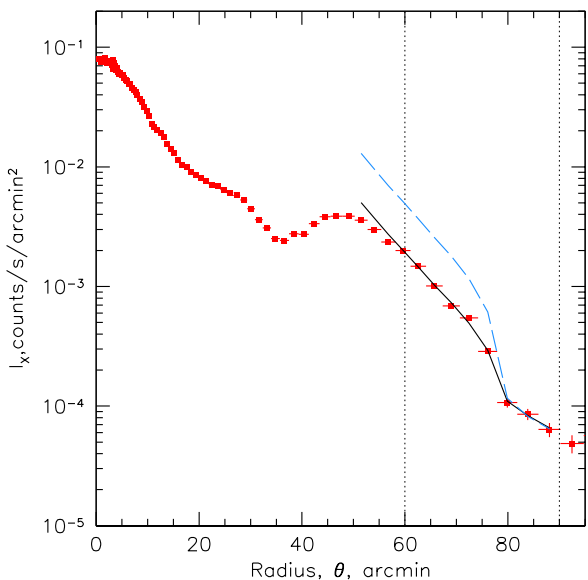}
\caption{Fit to the relic surface brightness profile with a broken power-law gas-density profile. The best-fitting edge position is at $79'$. At the edge the density changes by a factor of $\approx 2.18$, which in the gas with adiabatic index 5/3 corresponds to the shock Mach number $M\approx 1.9$. For comparison, the blue dashed curve shows the same model for $M=3.5$ that is needed to explain the slope of the relic integrated radio spectrum (under the standard set of assumptions). This model is clearly excluded by the data.
}
\label{fig:relic_shock}
\end{figure}

For a more quantitative characterization of the X-ray surface brightness edge in the relic region, we employed an often used spherical "jump" model to fit the profile \citep[see examples in][]{2007PhR...443....1M}. The model approximates the radial gas density distribution with two power laws below and above the  "shock" radius $r_s$, namely, 
\begin{equation}
n(r)=\left \{ \begin{matrix}
n_1r^{-a_1} & r<r_s \\
n_2r^{-a_2} & r>r_s 
\end{matrix} \right . 
.\end{equation}
The model makes a standard assumption of spherical symmetry (within the wedge) and of the X-ray flux in the 0.4-2.3~keV (or similar) energy band being weakly sensitive to the gas temperature. With these assumptions, the ratio of the best-fitting normalizations, $n_1$ and $n_2$, yields the density compression factor, which can be used to infer the shock Mach number. For the particular case of the Coma relic, the assumptions of spherical symmetry and the power law distributions are the strong ones. Nevertheless, we have formally applied this model to the relic wedge for the range of radii between $60'$ and $90'$ (see Fig.~\ref{fig:relic_shock}). The best-fitting values are as follows: $r_s=79'$, $\displaystyle \left(n_1/n_2\right)=2.18$, and $a_1=1.25$, $a_2=0.8$.  We do not quote uncertainties on the derived values since they are seriously affected by the choice of the radial interval for fitting and by the definition of the "relic wedge." The above value of the density jump corresponds to the shock with $M=1.9$ for the  Rankine-Hugoniot condition in the gas with the adiabatic index $\gamma=5/3$. For a spherically symmetric shock propagating in a power-law density profile, one would expect  the slope of the density profile downstream from the shock to be flatter than the initial value. Therefore, one can expect $a_1 < a_2$. This is not what is observed, most likely due to the overly simplistic model, given that the SW wedge has a very complicated structure.
In particular, a smooth transition from the region dominated by the gas of the NGC4839 group  to the region downstream of the shock casts doubts on the validity of a single power law model. Nevertheless, the derived values of $r_s$ and $M$ appear broadly consistent with the previous studies shown in Table~\ref{tab:mach}, once the differences in the choice of the center of the spherical shells are taken into account.

As discussed below, to explain the spectral index of the  relic integrated synchrotron emission $\alpha=1.18$ \citep{2003A&A...397...53T} in the frame of the standard diffusive shock acceleration (DSA) model with fast radiative losses, the Mach number $M\sim 3.5$ is needed.
The blue dashed line in Fig.~\ref{fig:relic_shock} shows the model with the same set of parameters as the best-fitting modes, except for the density jump, which now corresponds to $M=3.5$ and is clearly excluded by the data, at least for the assumed geometry of the shock front.

\begin{figure}
\includegraphics[trim=0cm 5cm 0cm 3cm,width=1\columnwidth]{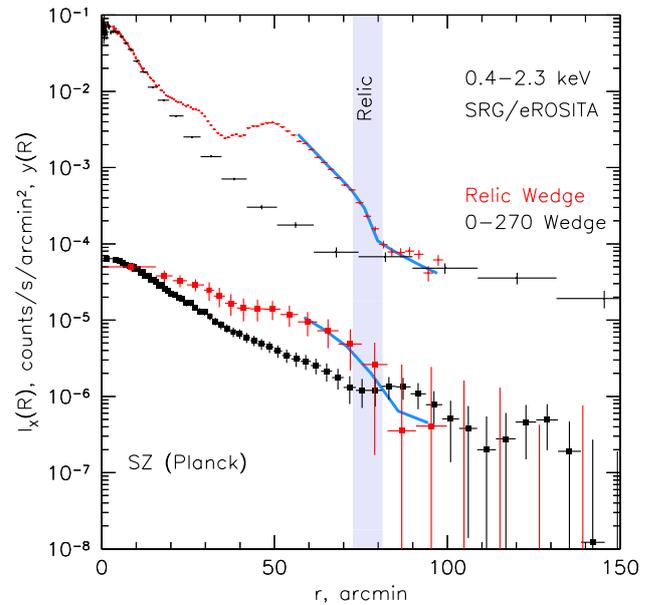}
\caption{Comparison of X-ray and SZ radial profiles in the relic (red) and non-relic (black) wedges. The blue line shows the shock model. For the tSZ data, the model was convolved with the FWHM=$10'$ Gaussian. Broadly, the tSZ profile in the relic wedge is consistent with the presence of the $M\sim 2$ shock as derived from the X-ray data.}
\label{fig:radial_sz}
\end{figure}

A comparison of the X-ray and tSZ radial profiles in two wedges is shown in Fig.~\ref{fig:radial_sz}. To generate the tSZ image, we used \textit{Planck}~ R4 HFI frequency maps \citep{2020A&A...643A..42P}. The all-sky maps were masked for bright sources and the Galactic Plane, while the resolution was degraded to $10'$ and a straightforward internal linear combination procedure (ILC) was applied, adding a constant as an additional map. For the ILC, the only requirement was the minimization of the derived tSZ scatter, while preserving its amplitude. The ILC coefficients have been calculated on the HEALPIX\footnote{ https://healpix.sourceforge.io} map subdivided into 192 patches. The resulting coefficients were then smoothed to cover the entire sky and the tSZ map was produced on a fine grid with the effective $10'$ resolution.  Figure~\ref{fig:radial_sz} shows that, broadly, the tSZ profile in the relic wedge is consistent with the presence of the $M\sim 2$ shock, although the constraints on the shock parameters are less tight than those based on the X-ray data.

\section{Spectra}
\label{sec:spec}
\begin{figure}
\includegraphics[trim=0cm 5cm 0cm 3cm,width=1\columnwidth]{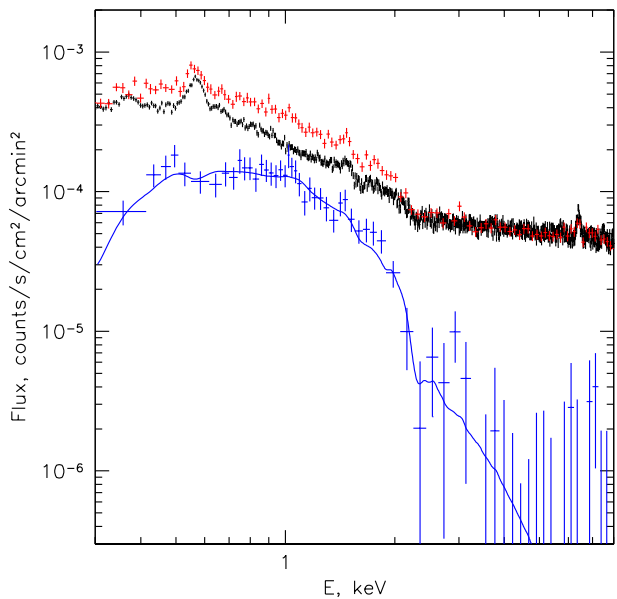}
\caption{Spectrum of the excess emission in the relic region (blue). This emission was calculated as the difference of the spectrum in the relic wedge between 70 and 80 arcminutes (red) and the spectrum in the same radial bin but in a wider wedge 0-270 degrees (black). The APEC fit to the relic emission is shown with a blue curve.
}
\label{fig:relic_spec}
\end{figure}

The correction of spectra for the stray light is considerably more complicated than the correction of the soft band images since it requires calibration of the stray light's energy-dependent radial profiles. To circumvent this problem, we simply extracted the spectrum in the $r=70-80'$ radial bin in the relic wedge and compared it with the spectrum in the non-relic wedge (for the same range of radii). The errors on the measured values were calculated following the recipe described in \cite{1996ApJ...471..673C}. To the first order, these spectra are expected to contain similar contributions of the stray light due to the bright Coma core. Corresponding spectra (red and black points) are shown in Fig.~\ref{fig:relic_spec}. The difference between these two spectra is shown as blue data points. We note in passing that even assuming that all backgrounds and foregrounds are cleanly subtracted, the difference between these two spectra,  corresponds to the difference between the emission coming from the volume filled with gas near the relic position and a similar volume in the non-relic wedge, that is, $s_{d}(E)=s_{r}(E)-s_{nr}(E)$, where $s_d$, $s_r$, and $s_{nr}$ stay for the difference, relic volume, and non-relic volume spectra, respectively.  

This difference spectrum can be reasonably well approximated by the APEC model. The abundance is poorly constrained by the data, except for an upper limit of $\sim0.15$ (for the adopted abundance table of \citealt{2003ApJ...591.1220L}).  The best-fitting temperature is $kT_e=2\pm 0.4~{\rm keV}$ and the absorbing column density, $N_H=(5.4\pm 1.6) 10^{20} \,{\rm cm^{-2}}$. This column density is higher than $N_H\sim 10^{20}\,{\rm cm^{-2}}$ estimated from HI surveys and the dust distribution in the Milky Way. However, it is plausible that the $s_{nr}(E)$ spectrum is softer than the relic spectrum, $s_r(E)$. As a result, in the difference spectrum, $s_d(E),$ the flux at the lowest energies might be over-subtracted, resulting in larger values of $N_H$ in the absorbed APEC model.
If we fix absorption to the Galactic value $N_H=9\times 10^{19} \,{\rm cm^{-2}}$, the best-fitting temperature goes up to  $kT_e=3.0\pm 0.7\,{\rm keV}$, consistent with previous studies. In any case, the measured temperature is significantly lower than expected for a $M=3.5$ shock. Indeed, assuming an upstream temperature of 1.5~keV \citep{2013PASJ...65...89A} and the temperature jump of $\approx 4.69$ for this Mach number, one gets $kT\approx 7\,{\rm keV}$ - significantly higher than the observed value. It is still possible that the flux from this region is dominated by cooler gas that originally belonged to the NGC4839 group, while the hot shocked gas occupies a very small volume. This is only possible if the geometry of the shock is far from the spherical model discussed above. However, the two-temperature fit carried out by \cite{2013PASJ...65...89A} did not reveal a very hot component in the relic region.


\begin{table}
  \caption{Mach number and the shock radius estimates for the Coma relic. For \textit{SRG}/eROSITA, quoted uncertainties correspond to estimated systematic errors, driven, in particular, by the freedom in choosing the radial range for fitting. The derived Mach number assumes that the curvature radius of the front is equal to the relic distance from the Coma center.} 
  \label{tab:mach}
  \centering
    \begin{tabular}{  lll }
    \hline
    Reference & Mach & $r_s$ [$'$] \\ \hline
    X-rays, \cite{2013PASJ...65...89A} & $2.2 \pm 0.5$ & -\\
    X-rays, \cite{2013MNRAS.433.1701O}  & $1.9^{+0.16}_{-0.4}$ & - \\
    SZ, \cite{2015MNRAS.447.2497E} & $2.9^{+0.8}_{-0.6}$ & $79^{+10}_{-9}$ \\
    SZ, \cite{2016AA...591A.142B} & $2.2\pm 0.3$ & $75$  (fixed) \\
    X-rays, \textit{SRG}/eROSITA & $1.9\pm 0.2$ & $79\pm 2$ \\
    \hline
    \end{tabular}
\end{table}

\section{Discussion}
\label{sec:dis}
The synchrotron emission of the Coma relic has been the subject of many studies \citep[e.g.,][]{1990AJ.....99.1381V,1991A&A...252..528G,2001A&A...366...26E,2011MNRAS.412....2B},  most recently published by \citealt{2022ApJ...933..218B} (see references therein) using the LOFAR data. Here, we briefly discuss two issues: (i) the morphology of the merger and of the shock and (ii) a notorious problem of a mismatch between the Mach numbers estimated from X-ray and radio data. We refer to the aforementioned studies for a more extended analysis of the radio properties of the relic.

\subsection{Merger geometry and NGC4789}
The morphology of the perturbed gas of the NGC4839 group and its low velocity with respect to the Coma cluster led to the suggestion that the group came from the NE side, crossed the Coma core, and is now close to the apocenter \citep[see, e.g.,][]{1994ApJ...427L..87B,2019MNRAS.485.2922L,2019ApJ...874..112S,2019MNRAS.488.5259Z}. In this model, the merger shock (driven by the NGC4839 group during its first infall) is now located somewhere toward the SW from the group. The relic, therefore, naturally fits into the merger scenario. The region near the relic is now well covered in X-ray, radio, and microwave bands and so, we can probe more subtle features that are not captured by the baseline model. 

Yet another consequence of the merger is the possibility of accelerating some of the galaxies that were initially lagging behind the NGC4839 group to velocities higher than the group velocity \citep[e.g.,][]{2014JCAP...06..057C,2021MNRAS.506..839Z}. As a result, these galaxies might be found well ahead of the group after the first passage through the Coma core. When the group is near its apocenter, these galaxies may still be moving with high velocities along the general direction of the merger. It is, therefore, plausible that the radio galaxy NGC4789, which is located further away from the Coma center than the relic, is one of such "accelerated" galaxies that managed to overtake not only the NGC4839 group -- but, also, the merger shock driven by the group. A similar configuration is also considered in \cite{2022ApJ...933..218B}. Since the shape of radio tails of  narrow-angle tailed (NAT) galaxies, such as NGC4789, depends on the ram pressure, it is interesting to estimate the change of the ram pressure upstream and downstream of the shock. In the merger-shock scenario, NGC4789 and the gas are moving in the same direction (away from the Coma center) before crossing the shock. Once the galaxy crosses the shock, the density goes down, but the velocity relative to the ambient gas goes up. It turns out that the ram pressure, namely, $\upsilon_{rel}^2\,\rho$, where $\rho$ is the intracluster medium (ICM) density and $\upsilon_{rel}$ is the galaxy velocity relative to the gas, increases after crossing the shock (see Fig.~\ref{fig:ramp}) if the galaxy velocity is not very high. In the opposite case, the ram pressure decreases instead. The implication is that the tail might be wider (or narrow) before crossing the shock, as sketched in Fig.~\ref{fig:ngc4789}, where we depict the case when the galaxy velocity is relatively low (and only slightly higher than the shock velocity) and, therefore, in the past, the ram pressure was lower. As a consequence, a wide area along the trajectory of NGC4789 could be enriched with relativistic particles (assuming that the jet's power was the same as it is now). In any case, in this scenario, the current position of the gas that was at the shock front  when NGC4789 was crossing it is somewhere downstream of the current shock location. It would therefore be interesting to search for signatures of the shock-crossing in
the radio emission from this region. These signatures could manifest themselves as abrupt changes of the radio emission, for instance, an edge in the radio surface brightness -- unless the spatial diffusion is rapid enough to erase the substructure. We did not find any compelling evidence for such signatures in the publicly available LOFAR data and leave the question for further studies.  In any case, these arguments only strengthen the possible role of  NGC4789, which was suggested long ago as the source of relativistic electrons for the relic \citep[e.g.,][]{1991A&A...252..528G,1998A&A...332..395E,2022ApJ...933..218B}.

One interesting implication of the above scenario is that it helps to reveal the merger direction with respect to the line of sight. Since (i) NGC4879 has higher recession velocity $\sim 8250\,{\rm km\,s^{-1}}$ than Coma ($\sim 6930\,{\rm km\,s^{-1}}$) and (ii) accelerated galaxies  predominantly follow the merger direction \citep[e.g.,][see their fig.4 and 7]{2021MNRAS.506..839Z}, we conclude that the NGC4839 group is likely further away from us than Coma. This means that when NGC4839 was crossing Coma from NE to SW, it was receding from us; that is to say that the NE part of the trajectory was closer to us than the SW part now. By extension, the filament also shares the same orientation.

\begin{figure*}
\centering
\includegraphics[angle=0,trim= 0mm 0cm 0mm 0cm,width=1.9\columnwidth]{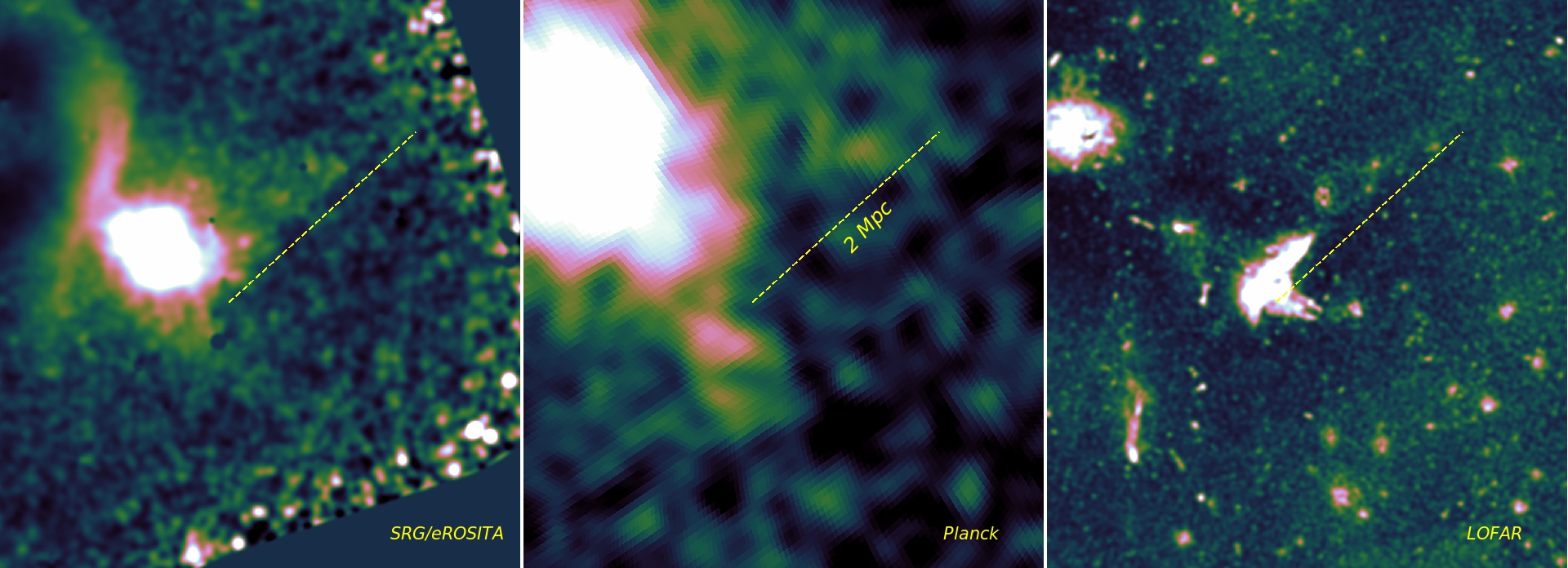}
\caption{X-ray, tSZ, and radio images of the relic region. The same dashed line in all plots has a length of 2~Mpc. The line is slightly shifted from the relic outer edge and the shock front for clarity. The relic outer edge appears to be aligned with the surface brightness edges in X-rays and tSZ (although the images, unlike the radial profiles, are too noisy to draw a firm conclusion). Beyond the brightest portion of the relic, both tSZ and X-rays show a faint extended emission and edge to the NW, approximately aligned with the relic edge. If this is a continuation of the shock, then its shape is far from a spherical surface centered at Coma. For a discussion of radio emission in this region, which is affected by the bright source Coma-A/NGC4827, we refer, for instance, to \citet{2011MNRAS.412....2B,2022ApJ...933..218B}.
On the SW side, an extension in the radial direction is seen in the tSZ and X-ray images, hinting at the presence of a gaseous filament along the direction towards A1367. }
\label{fig:xyr}
\end{figure*}

\begin{figure}
\includegraphics[trim=0cm 5cm 0cm 3cm,width=1\columnwidth]{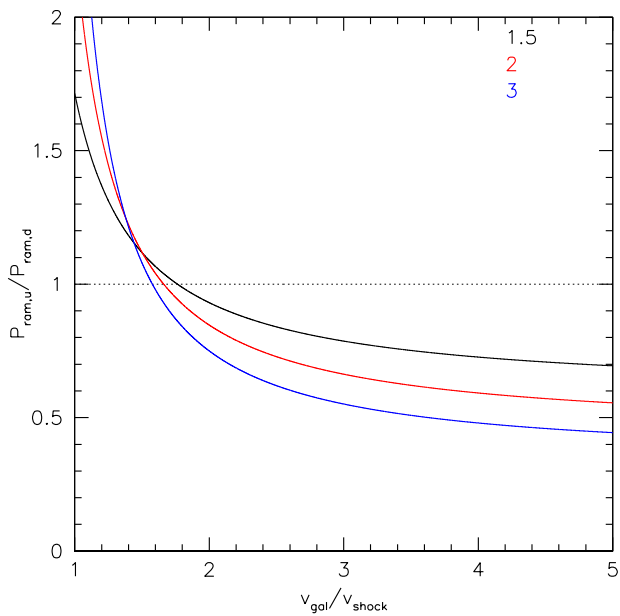}
\caption{Ratio of ram pressures upstream and downstream to the shock for a galaxy moving in the same direction, but faster than the shock. For the galaxy velocity smaller than $\sim 1.5$ times the shock velocity, the ram pressure increases when the galaxy takes over the shock (here, the upstream gas is assumed to be at rest). For a radio-galaxy with a bent tail, this means that the tail becomes narrower. For higher velocities, the tail becomes wider instead. The three curves represent  $M=1.5, 2, 3$ (black, red, and blue, respectively).}
\label{fig:ramp}
\end{figure}

\begin{figure}
\includegraphics[trim=0cm 0cm 0cm 0cm,width=0.8\columnwidth]{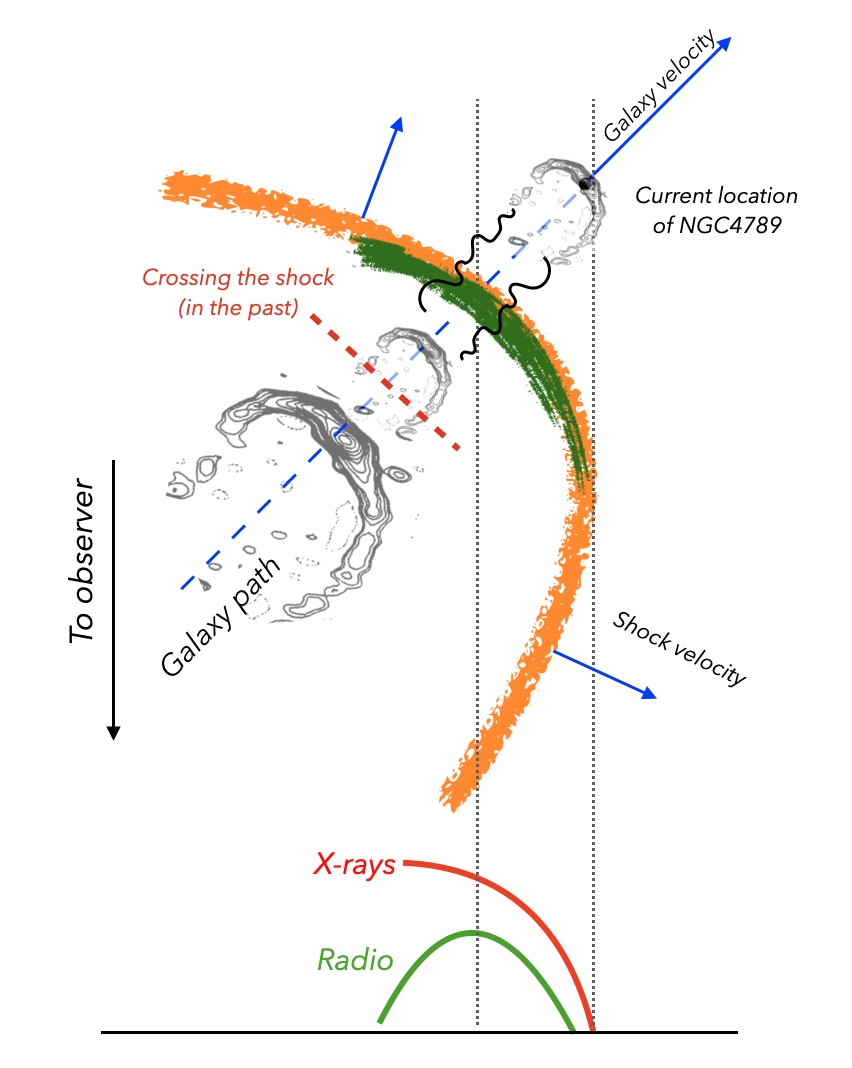}
\caption{NGC4789 (sketch, top view) in a pure merger shock scenario. The geometry is reminiscent of the one in
\citealt{1998A&A...332..395E} (see their Fig. 2), although the origin of the sketched configuration is different. Here, NGC4789 moves faster than the shock and eventually overtakes it. The ICM along the path of NGC4789 can be enriched by relativistic particles. The shock crossing happened in the past. We can hope to identify some changes in radio properties (e.g., the width of the enriched region) somewhere downstream from the current position of the shock (red dashed line). The radio emission at the shock reflects a more recent history of the system. 
As demonstrated by the numerical simulations of \cite{2019ApJ...876..154N}, when the galaxy moves away from the shock, the material along bent jets moves towards the shock and can even cross it. This is depicted by the black wave curves. 
}
\label{fig:ngc4789} 
\end{figure}

\subsection{Shock morphology}
\label{ssec:shock}
The presence of the steep gradient in the X-ray surface brightness and tSZ signal at the outer edge of the relic provides strong evidence for a shock in agreement with previous studies. In the idealized simulations considered in \cite{2019MNRAS.485.2922L,2019MNRAS.488.5259Z}, the shock associated with the NGC4839 group appears as a convex structure, enveloping the group and located further away from the Coma center, which is broadly consistent with the location of the relic. However, real data reveal a considerably more rich and more complicated structure. A zoomed-in view of the relic region in X-rays, tSZ, and radio bands is shown in Fig.~\ref{fig:xyr}. The brightest part of the radio relic has the size of $\sim 1$~Mpc and presumably traces the shock front. We note in passing that the radio relic does not necessarily trace the entire extent of the shock. For example, if the major contribution to the radio flux comes from the compression and re-acceleration of particles supplied by NGC4789, then only the part of the compressed gas enriched with such particles is going to be particularly bright (see, however, the discussion in \S\ref{sec:lowbeta} below). In contrast, in X-rays and tSZ, the shock should be visible (provided sufficient sensitivity) over the entire tangential (to the line of sight) surface of the shock. However, no obvious convex structure is seen in the X-ray or tSZ. If anything, there is either a small straight segment (approximately co-spatial with the radio relic)  
or a fainter concave structure extending to the NW and SW beyond the edges of the relic.

One possible explanation for this is that projection effects play a role here, emphasizing one particular part of the shock. Indeed, a clean convex shape is expected if the merger is happening largely in the sky plane, while for other projections, it may look more complicated, (see, e.g., Fig.B2 in \citealt{2021MNRAS.506..839Z} and  \citealt{2019ApJ...874..112S}). 

\begin{figure}
\includegraphics[trim=0cm 0cm 4cm 0cm,width=0.8\columnwidth]{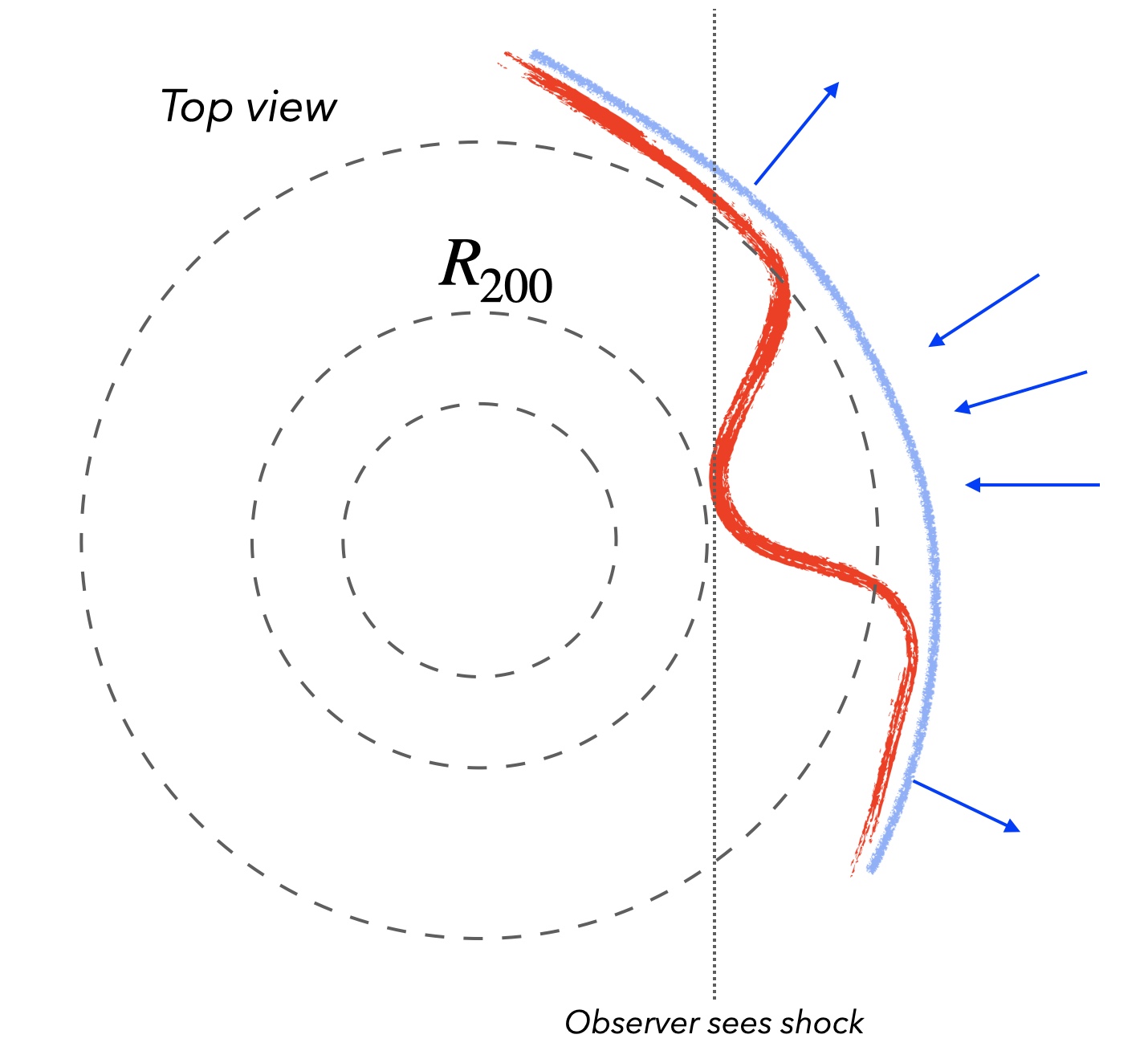}
\caption{Sketch of the shock geometry (a view from the top). The blue line shows the shape of an approximately spherical shock propagating through a hydrostatic atmosphere. This geometrical setup is used to derive the density jump of the X-ray emitting gas and to estimate the shock Mach number. The red line shows the shock modified by residual motions of the gas infalling along the filament. The curvature was intentionally exaggerated for clarity.}
\label{fig:shock_sketch} 
\end{figure}

Another, perhaps more plausible, explanation is the interaction of the outgoing merger shock with the infalling material along the filament connecting Coma with A1367, that is, along the SW-NE direction. In a self-similar collapse model \citep{1984ApJ...281....1F,1985ApJS...58...39B}, the position of the accretion shock is a function of the mass accretion rate (MAR), which is set by the logarithmic derivative of the cluster mass over the scale factor, $a,$ of the expanding Universe. Using the analytic approximation \citep{2016MNRAS.459.3711S} of the shock radius, specialized for $\Lambda$CDM and modest values of MAR, we can expect the accretion shock radius, $r_s\sim 2 r_{200c}\approx 140'$, that is, twice as far away from the Coma core than the relic. Of course, this applies only to a spherically symmetric case. Along the filaments, the shock can be  closer to the cluster core.  In fact, the accretion shock and its closer-to-the-center version ("infall shock") have been suggested as the origin of the Coma relic \citep[e.g.,][]{1998A&A...332..395E,2011MNRAS.412....2B}. 
If the accretion shock is that close to the Coma center, then the "collision" of the merger shock (driven by NGC4839) and the accretion shocks would create three discontinuities -- namely, two shocks and a contact discontinuity \citep[e.g.,][]{2003MNRAS.345..349B,2020MNRAS.494.4539Z}. We do not see any clear evidence for those structures in the data, although it is difficult to exclude this possibility. 
Even if the accretion shock is farther away from the Coma center than the relic, and the merger shock is solely responsible for the X-ray surface brightness jump, it is still possible that the shape of the merger shock has been modified by the momentum  of the infalling gas stream \citep[see, e.g.,][]{2016MNRAS.461..412Z}. In this case, the original convex shape of the outgoing shock can be bent inwards as shown in Fig.~\ref{fig:shock_sketch}. The shape of the shock front can be far from spherical and in the sky plane the tangential to the shock surface can be either convex or concave and its apparent size can be small. This might have a serious impact on the derived Mach number since the observed jump in the surface brightness was converted to the compression ratio assuming that the curvature radius corresponds to the distance of the shock front from the Coma center. If the curvature radius is smaller than this, the true Mach number will be larger than the value given in Table~\ref{tab:mach}. 
Indeed, the X-ray profiles measure an excess of the surface brightness, which is $\propto l\times n^2$, where $n$ is the gas density and $l$ is the assumed length along the line of sight. The density jumps ($\gamma=5/3$ gas) for $M=3.5$ and $M=1.9$ are $3.2$ and $2.2$, respectively. Therefore, $l$ should be a factor of $(2.2/3.2)^2\sim 0.5$ smaller than assumed in the model (see also the discussion at the end of \S\ref{sec:spec}). Close to a spherical shock with a curvature radius, $R$, the length of the line of sight going through the shocked gas is $l=2\sqrt{2R\times dx}$, where $dx$ is the distance from the shock front. Therefore, we can imagine that $R\sim 0.5-1\,{\rm Mpc}$ will be needed to make a $M=3.5$ shock appear as the $M=1.9$ jump in the X-ray surface brightness profile. Yet another consequence of a non-spherical geometry of the shock is a likely variation of the Mach number along the front, which means that the highest Mach numbers might not be present at the part of the shock surface seen by an observer as a surface brightness jump. For instance, for a bow shock not in the plane of the sky (Fig.~\ref{fig:ngc4789}), the apparent Mach number can easily be underestimated \citep[see, e.g., fig.11 in][]{ 2019MNRAS.482...20Z}. In this case, however, we need a mechanism for transporting relativistic electrons from the most efficient acceleration location (highest Mach number region) to the apparent shock edge to have the radio and X-ray upturns at the same place.   We return to this issue in Section~\ref{sec:lowbeta}.

In summary, we conclude that the evidence for the shock is strong, but its complicated morphology precludes an accurate determination of the shock parameters. Clearly, deeper dedicated observations of this region are needed to unambiguously determine the nature of the edges beyond the nominal extent of the relic. This applies not only to the X-ray band but to the SZ and radio bands, too. In the radio band, some extended structures are seen \citep[][]{2011MNRAS.412....2B,2022ApJ...933..218B}, although they 
might be affected by residuals remaining after the subtraction of the bright radio source Coma~A.

\begin{figure}
\includegraphics[trim=0cm 5cm 0cm 3cm,width=1\columnwidth]{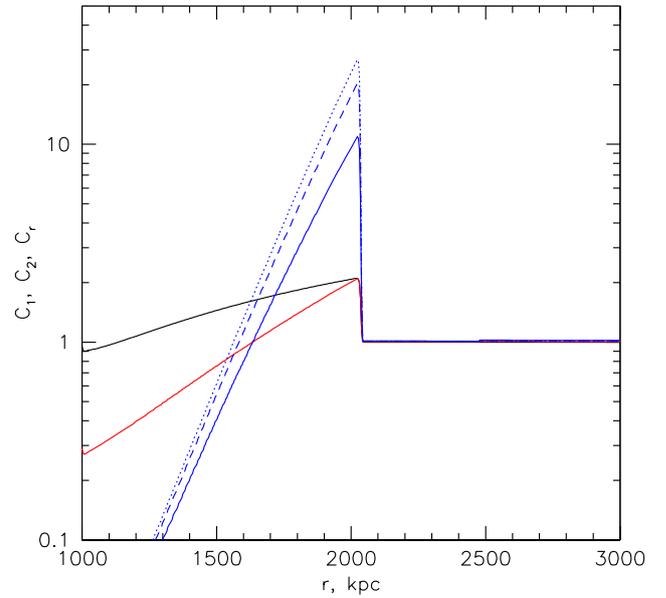}
\caption{Adiabatic compression and expansion scenario (van der Laan model). Three curves show different compression factors. The black curve shows the ratio of the current density at a given radius to the initial density at this radius, i.e., $C_1(r,t)=\frac{\rho(r,t)}{\rho(r,t=0)}$. Currently, the shock is at $r=2000\,{\rm kpc}$.
The red curve shows the ratio of the current density at a given radius to the initial density of the same gas lump, i.e., $C_2(r,t)=\frac{\rho(r,t)}{\rho(r[t=0],t=0)}$. These two factors differ because the shock displaces the gas from its initial location. 
The solid blue curve shows the radio emissivity boost factor in the van der Laan model for $p=3.36$. This model assumes pure adiabatic evolution of the relativistic particle spectrum in response to the compression of the gas ($=C_2$). The blue solid line corresponds to the case when  thermal and non-thermal plasmas are mixed on microscopic scales and the contribution of non-thermal particles to the total energy density is subdominant. For comparison, the dotted and dashed blue curves correspond to the adiabatic compression of a volume-filling relativistic plasma (dotted) and to the relativistic plasma confined to small bubbles and filaments (dashed). The observed width of the bright part of the radio relic is $\sim 3'$, which is $\lesssim 100$~kpc, corresponding to $r\approx 2000-100 \approx 1900\,{\rm kpc}$. The shock was passing through the gas at this location approximately 100~Myr ago. This suggests that adiabatic losses have a relatively minor impact on the radio emission on these spatial scales -- unless the spectrum is very steep.  
}
\label{fig:relic_c2prof}
\end{figure}

\begin{figure}
\includegraphics[trim=0cm 5cm 0cm 3cm,width=1\columnwidth]{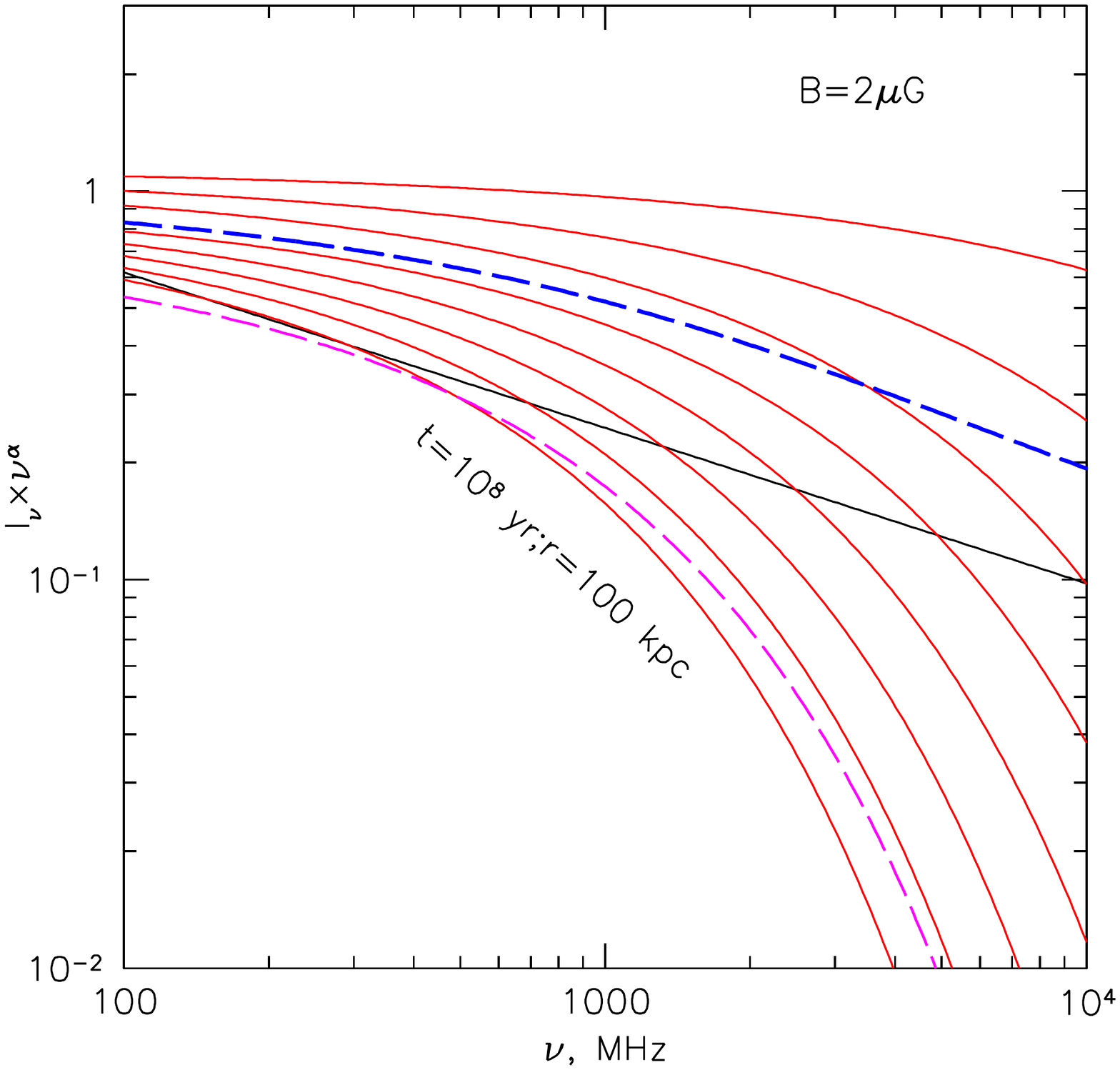}
\caption{
Spectra for $B=2 {\rm \mu G}$, near the shock and at several positions  downstream of the shock (up to 100 kpc), normalized by the spectrum at the shock. In terms of time, the red curves are "spaced out" by $\approx 13\,{\rm Myr}$. Both the radiative losses and adiabatic expansion  are taken into account. The magenta dashed line shows the spectrum at 100 kpc from the shock if changes in the particles' Lorentz factor and magnetic fields associated with the re-expansion of the gas are ignored. The blue dashed line shows the integrated spectrum over the regions up to 100~kpc from the shock. The thin black line shows the power law with the index large by 0.4 than the initial index at the shock front \citep[see Fig.19 in][]{2022ApJ...933..218B}.
}
\label{fig:lifemax}
\end{figure}

%
\subsection{Radio emission}
Several models explaining the emission of radio relics have been suggested thus far \citep[see, e.g.,][for reviews]{2014IJMPD..2330007B,2019SSRv..215...16V}. As the rule, they involve shocks and focus on various acceleration mechanisms of electrons either from the thermal pool or from the pre-existing non-thermal population. The radio emission from the Coma relic has been extensively discussed in the literature, most recently in \cite{2022ApJ...933..218B}. Here, we do not go into the detailed modeling, but briefly discuss a few aspects pertinent to the observations described in the previous sections.

\subsubsection{Radio and X-ray Mach numbers}


In the DSA theory, the spectral index, $p,$ of the accelerated particle distribution (per unit energy) at the shock front is linked to the shock compression ratio $C$ \citep[e.g.,][]{BE87}. The corresponding slope of the synchrotron emission is $\alpha=(p-1)/2$. The slope $\displaystyle p_s = \frac{C+2}{C-1}$ is established if accelerated particles are directly injected by the shock from a thermal pool. If there is a balance between the acceleration of new particles and the aging of particles via radiative losses, the slope of the integrated distribution of particles will be steeper by $1$, namely, $p_i=p_s+1$. The value of the shock compression ratio, $C,$ can be further linked to the Mach number (for a given adiabatic index, $\Gamma,$ of the medium) and can thus provide the relation between the Mach number, $M,$ and the spectral index, $\alpha$. In particular, for $\Gamma=5/3$, the Mach number is related to the synchrotron emission spectral indices $\alpha_s=(p_s-1)/2$ and $\alpha_i=(p_i-1)/2=\alpha_s+0.5$ as:
\begin{eqnarray}
M^2&=&\frac{(2\alpha_s+3)}{(2\alpha_s-1)} \label{eq:ms},\\
M^2&=&\frac{(\alpha_i+1)}{(\alpha_i-1)} \label{eq:mi}.
\end{eqnarray}

We note that the relations represented by Eqs. (\ref{eq:ms}) and (\ref{eq:mi}) are approximate and they are valid when the particle scattering centers that provide the diffusive shock acceleration are simply advected with the bulk plasma speed. However, if the DSA scattering centers have a sizable drift velocity relative to the bulk plasma (e.g., the Alfvenic drift velocity), then we should distinguish between the shock compression ratios of the bulk plasma and that of the scattering centers. The DSA particle spectra depend on the scattering centers compression ratio, $C_{sc}$ \citep[see e.g.][]{Bell78,KangRyu18}.  In a simple test particle case, the scattering center compression ratio is given by $\displaystyle C_{sc} = \frac{u_1 + U_{w1}}{u_2 + U_{w2}}$, where $U_{wi}, (i=1,2)$ are the mean velocities of the scattering centers in the rest frames of the upstream and downstream plasma flows, respectively. As we discuss below, the shock might be corrugated on relatively small scales due to the presence of non-thermal filaments.  Strong amplification of the magnetic field and wave-vector anisotropy can provide values of $U_{w2}$ that are large enough to be comparable with the downstream velocity in the rest frame of the shock, $u_{2}$. It appears that the effects of the non-zero drift of the scattering centers in DSA could affect the value of $M_R$ via a modification of Eq.~\ref{eq:mi}; however, rather special conditions are needed to make $M_R$ (for a given slope $\alpha$) smaller than what follows from the above equations.

\subsubsection{Mach number discrepancy and fine structure of shocks}
As already mentioned in \S\ref{sec:radial}, the observed integrated spectral index of the Coma relic is $\alpha=1.18$ \citep{2003A&A...397...53T} and from Eq.~\ref{eq:mi}, the expected Mach number is $M_R\approx 3.5$, which is significantly larger than the Mach number derived from the X-ray radial profile, $M_X\approx 1.9$ (see also Table~\ref{tab:mach}). It is interesting to note that if Eq.~\ref{eq:ms} is used instead, then $\alpha=1.18$ corresponds to $M_R\approx 2$, meaning that it is very close to $M_X$. 

This discrepancy is also known for several other relics. Indeed, the 
analysis of the radio and X-ray observations of a sample of relics established that Mach numbers estimated from radio relics by using the test particle DSA model as the source of accelerated electrons are, on average, higher than the Mach numbers estimated from the X-ray data \citep[e.g.,][and references therein]{2017A&A...600A.100A,2020A&A...634A..64B,2021MNRAS.500..795D,Wittor21MN,2021arXiv211200023W}. In some of these relics, strong support for interpreting $\alpha$ as the integrated spectral index of aging electrons comes from spatially resolved data that show progressive steepening of the spectrum on the downstream side and a power slope of the integrated spectrum. The Toothbrush relic is a good example that features a power law shape of the synchrotron emission of the integrated spectrum over almost three decades in frequency \citep[e.g.,][]{2020A&A...642L..13R}.  Unless these particles already have sufficiently hard spectra, high Mach numbers are still needed in the DSA model to get the observed integrated radio spectral index, which might only be present in a fraction of the shock surface  \citep[e.g.,][]{2020A&A...634A..64B}.  Based on high-resolution cosmological simulations, \citet[][]{Wittor21MN} generated X-ray and radio
mock observations of shock fronts in merging clusters, also taking into account projection effects (see also \citealt{2015ApJ...812...49H}). They found similar behavior in the mock data and concluded that while 
the synchrotron radio emission is dominated by the high Mach number regions of the shock, the brighter X-ray emission is connected with the lower Mach number portions of the shock. Another problem associated with low Mach numbers is the requirement of having extremely high efficiency of converting the shock kinetic energy flux into the energy of accelerated particles needed to power the observed radio flux,  which  can be alleviated if relativistic particles are already present upstream of the shock \citep[e.g.,][]{2011ApJ...728...82M,2020A&A...634A..64B}.

In principle, this scenario (pre-existing particles, broad distribution of Mach numbers, and projection effects) can be  applied to the Coma relic in a straightforward way, whereby NGC4789 can supply the relativistic particles upstream of the shock. We add here that the shock can have  a complicated substructure due to the presence of non-thermal filaments associated with radio galaxies. The difference in sound speeds between thermal and non-thermal phases can make the shock front strongly corrugated and produce extra weak shocks as well as turbulence both upstream and downstream of the shock \citep[e.g.,][]{2005ApJ...634L.141H,2012ApJ...746..112F}. These effects might contribute to the production of turbulence and scattering centers in plasma and to the breadth of the Mach number distribution.

\subsubsection{Low $\beta$ filaments and fast propagation of relativistic electrons}
\label{sec:lowbeta}
Going back to the example of the Toothbrush relic, a striking feature of its radio spectra is the similarity of the integrated spectral slope at different locations along the full extent of the relic \citep[e.g.,][]{2020A&A...636A..30R}. While this stability of the slope might mean that the Mach number distributions in various locations are similar (with the high end being the most significant), yet another interesting possibility is the fast propagation of electrons along the relic. This would imply that the integrated slope reflects the acceleration efficiency and the Mach number distribution in the entire relic, rather than locally. Variations of the spectral slope across the relic (progressive steepening with the distance from the outer edge) show that the propagation of particles in this direction is suppressed. The requisite conditions for displaying such behavior could be provided by a pre-existing filament (or a bundle of filaments or a sheet) with low plasma beta $\beta_{\rm pl}$ that is crossed by a shock. We may consider, for example, a filament dominated by non-thermal particles and magnetic fields, which could represent an evolved bubble of relativistic plasma initially inflated by AGN and then deformed and stretched by ICM motions \citep[e.g.,][]{2021ApJ...914...73Z} or a segment of a radio tail. Given the proximity of NGC4789, this could plausibly be the source of these filaments. We assume that inside the filament the magnetic and non-thermal particle energy densities are comparable and their sum balances the ICM thermal pressure outside the filament; therefore, $\beta_{\rm pl}=8\pi P_{f}/B^2_{f}\sim 8\pi P_{ICM}/B^2_{f}\sim 1$, where subscript $f$ is used for quantities inside the filament. Such structures could be analogs of loops and filaments that are discussed in \cite{2021NatAs...5.1261B,2022ApJ...934...49G,2022ApJ...935..168R}. On the one hand, this implies that the Alfven velocity inside the filaments is very high (much higher than the ICM sound speed) and, on the other hand, that these filaments are resilient to small-scale bending by the turbulent motions in the ambient ICM and are not subject to the mirror or firehose instabilities. In the outskirts of clusters, at radii $\sim R_{500}$ -- $R_{200}$, the energy density of the thermal gas is comparable to the CMB energy density. Therefore, inside the magnetic-field-dominated filament, the lifetime of electrons emitting synchrotron radiation at a given frequency is close to the longest possible. In contrast, in a filament located in the cluster core,  the lifetime of particles would be much shorter. We speculate that in the Coma case, the presence of such a structure means that the DSA scenario might still be consistent with the current data if the highest Mach number occurs not at the tangential surface seen in X-rays as a shock front -- but somewhere else (see Fig.~\ref{fig:ngc4789}) and electrons are spread along the routes provided by filaments and sheets. Similarly, in other relics (including Toothbrush), this process might explain the peculiar morphology of radio emission that cannot be easily explained via the surface of the shock front. 
We defer a detailed discussion of this model to a future work.

\subsubsection{Compression and expansion scenario and non-radiative losses}
\label{sec:nrlosses}
Locally, aging leptons from the tails of radio-galaxies can be re-accelerated by shocks as well \citep[see e.g.,][]{tail_galaxies17}. Here, NGC4789 is the prime candidate \citep[e.g.,][]{1991A&A...252..528G}. The spectrum of pre-existing relativistic electrons $p_0$ might be harder than the DSA prediction, $p_s$ for low $M$, namely, $p_s>p_0$.  Then the particle distribution at the shock will have a power-law index of $p=\min{(p_s,p_0)}$. This implies that the low Mach-number shocks will boost the synchrotron volume emissivity, but keep the spectral slope of the initial non-thermal particle population, if there are no patches with high Mach numbers that generate $p_s>p_0$ \citep[][]{2020A&A...634A..64B,Wittor21MN}. Even without re-acceleration, purely adiabatic compression at the shock will boost  synchrotron emissivity, while preserving the slope of the spectrum. In this subsection, we briefly discuss the scenario where compression and expansion play a major role.

Relativistic electrons suffer from radiative (synchrotron and inverse Compton) losses and are affected by adiabatic losses  (gains) due to the expansion (compression) of the gas. The radiative losses are controlled by the energy density of magnetic fields and radiation, namely, $B^2/(8\pi)+U_{CMB}$, where $U_{CMB}$ is the energy density of the cosmic microwave background. The ICM is commonly described as a high $\beta$, weakly collisional plasma, where $\beta=8\pi P/B^2$ and $P$ is the gas pressure. For estimates, we use the value of the gas pressure upstream of the shock $P_0\sim 1.5\,10^{-4}\,{\rm keV\,cm^{-3}}$ by extrapolating the radial profiles  from \cite{2013ApJ...775....4S,2020MNRAS.497.3204M} to $80'$. For $M=1.9$ shock, the pressure jump is $j_p\sim 4.3$. Therefore, on the downstream side of the shock, $\beta=100$ and $1$ would correspond to $B = (8\pi j_p/\beta) ^{1/2}\sim 0.5$ and $\sim 5\,{\rm \mu G}$, respectively. We assume that the magnetic field is somewhere in this range. 

We now consider adiabatic losses in the frame of a spherical shock model with $M=1.9$ propagating through a hydrostatic gaseous atmosphere approximately matching the density profile of the Coma cluster. We first consider the simplest case of a uniform magnetic field and assess the roles of tadiabatic expansion and radiative losses. Simultaneously, we consider the magnitude of the synchrotron emission boost factor that is driven by the compression of the existing population of relativistic particles in the ICM by the shock without any acceleration or re-acceleration. Such adiabatic compression by shocks has been considered as one of the possible mechanisms of reviving fossil radio plasma in clusters \citep[e.g.,][]{2001A&A...366...26E,2019MNRAS.488.5259Z}. For supernova remnants, such a model has been developed by van der Laan \citep{1962MNRAS.124..125V}. As in \cite{2019MNRAS.488.5259Z}, we plot (in Fig.~\ref{fig:relic_c2prof}) the boost factor of the radio emissivity that is expected in a pure adiabatic compression of pre-existing particles with the power law slope of $p=3.36$.  The three curves in Fig. \ref{fig:relic_c2prof} show the change of the ISM density at a given radius $C_1=\rho(r,t)/\rho(r,t=0)$; the density ratio of the gas at a given radius to the initial density of the same gas lump $C_2=\rho(r,t)/\rho(r(t=0),t=0)$; and, finally, the boost in the volume emissivity in radio band $C_r$, which in the van der Laan model is $C_r=C_2^{2/3p+1}$, provided that magnetic field changes across the shock by a factor $C_2^{2/3}$ (note:\ for a perpendicular shock, this factor would be simply $C_2$). Three variants of the boost factor $C_r$ are shown: the blue solid line corresponds to the case when  thermal and non-thermal plasmas are mixed on microscopic scales and the contribution of non-thermal particles to the total energy density is subdominant. For comparison, the dotted and dashed blue curves correspond to the adiabatic compression of a volume-filling relativistic plasma (dotted) and to the relativistic plasma confined to small bubbles and filaments (dashed), respectively. This parameterization ignores radiative losses completely but illustrates the impact of the expansion or compression on the radio flux. Given that the width of the bright part of the radio relic is $\sim 1.5-3'$, namely, less than 100~kpc, it is clear that over this distance, the expansion of the gas has a minor impact on the radio flux; however, on a distance that is a few times greater,  the expansion seriously affects the emissivity. For instance, $\sim 400\,{\rm kpc}$ downstream, the volume emissivity decreases by an order of magnitude purely due to the expansion. If the curvature radius of the shock is significantly smaller than used to model the density jump, then the adiabatic losses might be important at even smaller distances from the shock. As mentioned previously, the Mach number can be underestimated as well. A higher Mach number would imply a harder initial spectrum (in the DSA model), while the more rapid adiabatic losses can affect the relation between the slope of the integrated relic spectrum with the initial one. Therefore, these two effects are working in the same direction and, in principle, might reduce the tension between the X-ray and radio data.  

Next, we consider the case when the radiative losses are minimal under the assumption of a uniform magnetic field. The longest radiative life-time of relativistic electrons emitting at a given frequency is achieved when $B_1=\left (8/3\,\pi\,U_{CMB} \right )^{1/2}\approx 2\,{\rm\mu G}$ at the Coma redshift. 
\cite{2022ApJ...933..218B} found steepening of the spectral index (between 144 and 326~MHz) by 0.4 across the relic - from $\alpha\approx 1.2$ at the relic outer edge to $\alpha\approx 1.6$ some $4'$ towards the inner edge. We therefore consider the aging of the spectrum over the region $\sim 100$~kpc downstream of the shock, assuming $B=B_1$ and taking into account the radiative and adiabatic losses (Fig.\ref{fig:lifemax}). This exercise confirms that: (i) the adiabatic effects (in a spherical shock scenario) are subdominant and (ii) on spatial scales $\sim 100$~kpc, the estimated steepening matches the observed one.\footnote{We note in passing that this calculation ignores projection effects, which can have a strong impact, especially in the re-acceleration scenario involving NGC4789. In reality, the thickness of the layer where significant steepening takes place can be much smaller.} 

To complete the discussion of the adiabatic compression or expansion scenario, we note that if NGC4789 is supplying relativistic particles to the relic, it is difficult to explain the relic emission by simple compression (without re-acceleration). Indeed, \cite{2022ApJ...933..218B} reported spectral index $\alpha\approx 1.6$ of the NGC4789 tails ahead of the relic. However, the relic spectrum has a harder slope $\sim 1.2$ up to 4-5~GHz \citep{2003A&A...397...53T}. Even assuming that there is a flattering of the tails' spectra just below the LOFAR frequency, in the  adiabatic compression scenario, the break frequency  has to move up by a factor of 30-40 to explain the relic spectrum up to 5~GHz. This is unlikely even in the most favorable compression scenario. We therefore\ believe that acceleration or re-acceleration is necessary. 

\subsubsection{Intermittent and decaying magnetic field}
Yet another possibility to reduce the role of radiative losses 
is the spatial intermittency of the magnetic field. Various scenarios have been considered, starting from a series of papers by Tribble \citep[e.g.,][]{Tribble93}.   Here, we refer to a situation where (i) $\langle B^2/8\pi\rangle$ is smaller than the CMB energy density $U_{CMB}$, whereas in a small fraction of volume, $B\sim B_{max}$ is typically much larger than $\langle B^2\rangle^{1/2}$ and (ii) all electrons are probing the entire range of $B$ variations on short time scales. In this case, most of the observed radiation could be due to $B \sim B_{max}$ regions, while the aging is set by $\langle B^2+U_{CMB}\rangle\approx U_{CMB}$. The steepness of the electron spectrum helps since the synchrotron emission is $\propto \langle B^{(p+1)/2}\rangle$ and for $p>3$, it can be dominated by the largest values of $B$, which do not contribute much to aging. Of course, for $p=3.36$, high values of $B_{max}$ are needed to ensure that the radio emission and aging of electrons are "decoupled." For the ultra-steep spectra, the role of this process might be much more important. 

With the "prolonged" aging of electrons, the adiabatic losses increase and the resulting integrated spectrum can have a slope  closer to the one given by Eq.~\ref{eq:ms}.  Alternatively, we can imagine that the magnetic field is strong close to the shock front, but quickly decays downstream on timescales shorter than the effective cooling time of electrons. Once again, this might shift the shape of the integrated spectrum closer to the initial spectrum downstream of the shock.

The two scenarios described in this section and in \S~\ref{sec:nrlosses} might partially alleviate the 
X-ray and radio Mach numbers discrepancy. However, simultaneously they exaggerate the acceleration efficiency problem since the significant increase of non-radiative losses has to be compensated. 

Along the same lines, we can imagine that the radio emission of the relic is not stationary. For instance, if the shock has just arrived at the region enriched with pre-existing relativistic electrons, then for $t\lesssim t_{cool}\sim 10^{8}\,{\rm yr}$, Eq.~\ref{eq:ms} would be more appropriate than Eq.~\ref{eq:mi}. Here, $t_{cool}$ is the radiative cooling time of electrons that generate the highest frequencies in the relevant range. For $t\ll t_{cool}$, the apparent transverse size of the relic would not reflect the thickness of a narrow shell but rather the projection of the radio-bright surface on the sky plane. If relativistic electrons are able to spread very quickly (see \S\ref{sec:lowbeta}), this does not create any additional fine-tuning problem (except for observing the relic at a "special" time). However, the acceleration efficiency remains a problem, since the same total flux has to be explained. Overall, it seems unlikely that these scenarios would work without further modifications, unless other, rather special conditions are present in the Coma relic.

\subsubsection{Hadronic scenarios}

Given the long lifetimes of relativistic protons in the intercluster medium \citep[][]{VAB96,BBP97}, they may experience acceleration by multiple weak shocks and large-scale MHD plasma motions at the cluster scale $r_{200}$ which can form a power-law distribution of GeV-TeV regime protons with  indices of  $p$ = 2.4 -- 3  \citep[see e.g.,][]{B19}. Then the diffuse electrons of the spectral index about 2.4-2.6 can be re-accelerated by the relic shock of $M_x \lesssim$ 2, providing the possibility to relax the apparent $M_X$-$M_R$ discrepancy.

The origin of at least a fraction of radio-emitting electrons in clusters and relics may be connected with secondary electrons produced by hadronic collisions of relativistic protons with the thermal ones \citep[see e.g.,][]{Dennison80,Blasi99,Dolag2000}. This process can be traced via gamma-ray observations of the clusters of galaxies. Recently, \citet{2021A&A...648A..60A,2022MNRAS.516..562B}  claimed a detection with Fermi-LAT of a diffuse GeV emission from Coma. In the hadronic model of the gamma-ray origin, this corresponds to proton spectra with a power law with a  spectral index of about 3. 
  
Within some simplified model  \citet{2021A&A...648A..60A} estimated that the synchrotron radio emission from the secondary electrons produced together with gamma-rays in the hadronic interactions is four to six times below the total cluster radio emission flux -- assuming a steady state. With the present level of uncertainties among the reported gamma-ray spectra and fluxes -- and while keeping in mind the intermittent character of the magnetic field in the Coma cluster -- it is not easy to draw a firm conclusion on the possible role of the re-acceleration of the secondary electrons to the radio emission of the localized structures such as the Coma relic. However, the low density of the
ICM in the cluster outskirts represents a challenge for this scenario,
since the "local" production rate of secondary electrons is likely to be low.

\vspace{1cm}

In summary, all scenarios considered above feature some inherent difficulties. The most straightforward version of DSA, with acceleration from the thermal pool and the assumption of a spherical shock with a curvature radius equal to the relic distance from the Coma center, is certainly not an adequate description of the relic region. A combination of  pre-existing particles (presumably from NGC4789), broad distribution of Mach numbers in the shock, and projection effects (that have been discussed for other relics) appears to be a viable scenario for Coma too. In addition, we speculate that a very rapid propagation of electrons along the relic and the role of relativistic filaments on the substructure of the shock might be 
important. We defer a more detailed discussion of these effects to future studies.

\section{Conclusions}
We analyzed \textit{SRG}/eROSITA X-ray observations of the Coma cluster, focusing on the radio relic region, drawing the following conclusions. 
\begin{itemize}
\item A clear edge is present in the Coma X-ray image almost co-spatial with an outer edge of the radio relic (\S\ref{sec:images}). Fitting the edge with a jump in the ICM density radial profile yields the density ratio at the jump $\sim 2.18$, corresponding to the Mach number $M\approx 1.9$ for the monoatomic non-relativistic gas (\S\ref{sec:radial}). This is in broad agreement with other X-ray-based results in the literature as well as with the Planck SZ data. \item The excess emission in the relic region (compared to emission in the same radial range but in other directions) can be approximated by the thermal emission of an optically thin plasma with $kT\sim 2\,{\rm keV}$ (\S\ref{sec:spec}). What is somewhat surprising is the lack of clear signs of the transition from the ICM associated with the NGC4839 group and the gas compressed by the relic shock in the X-ray surface brightness profile (see Fig.~\ref{fig:radial_relic}). 
\item While the size of the radio relic might be limited to the area where relativistic electrons are present (upstream from the shock), the X-ray edge should be visible over the entire extent of the shock (subject to projection effects). We do see a faint X-ray edge beyond the relic boundaries. There are intriguing hints of similar structures in radio and SZ signals, however deeper data in each band are needed to unambiguously determine the nature of these extended structures and their spatial correspondence (\S\ref{sec:dis}).
\item In the pure Coma+NGC4389 merger scenario, the shock driven by the NGC~4389 group is expected to have a rather simple, quasi-spherical shape and it is further away from the Coma than the group. While the location of the relic is qualitatively consistent with this scenario, the morphology of X-ray emission in this region appears to be much more complicated. We tentatively associate this complexity with the inherently perturbed state of the gas beyond $r_{200c}$ along the major filament in the Coma-A1367 direction. This implies that the spherical shock approximation may be too simplistic and the value of the Mach number derived from X-ray data might be biased toward the low end (\S\ref{ssec:shock}).    
\item It is plausible that the merger with NGC~4389 has accelerated the radio galaxy NGC4789 to such a high velocity that it overtook not only the NGC~4839 group but also the forward shock \citep[see also][for the discussion of a similar configuration]{2022ApJ...933..218B}. It is most likely that NGC~4789 had come from the same filament and was initially lagging behind the group. The implication is that downstream from the current position of the shock there must be a region that corresponds to a shock crossing by the galaxy in the past. The "memory" of the  crossing of the NGC~4389 group atmosphere by NGC~4789 might be reflected in the current properties of the region between these two galaxies. (\S\ref{ssec:shock})
\item Another implication of the above scenario is the information on the orientation of the merger direction (and, perhaps, of the major filament) with respect to the line of sight.  (\S\ref{ssec:shock})
\item Pure adiabatic compression (van der Laan model) appears to be insufficient to produce a $\alpha=1.2$ radio spectrum of the relic and, thus, an acceleration or re-acceleration of particles is needed.
\item As in several other relics, the Mach number derived from the X-ray data ($M_X\sim 1.9$) turns out to be lower than  that obtained from the radio spectra ($M_R\sim 3.5$) in the DSA framework, assuming that the radiative losses are dominant and fast. We briefly discuss the effects, which might bring these estimates closer to each other. 
\item A combination of  pre-existing particles (presumably from NGC4789) coupled with a broad distribution of Mach numbers actively discussed for other relics appears to be a viable scenario for the Coma relic as well.
\item In addition, we consider several other effects that might be relevant. In particular, the possibility of fast spreading (much faster than the ICM sound speed) of energized electrons in non-thermal low plasma beta  structures \S~\ref{sec:lowbeta}, which can bring electrons accelerated somewhere at the shock surface to the region that is observed in X-rays as the shock front. In other relics, this could naturally produce the same shape of the radio emission along the entire area occupied by such structures.

\item We argue that non-thermal filaments (parts of the NGC4789 tails) stretching across the shock can make the shock front corrugated, as well as promote turbulence and generate a network of weak shocks on both sides of the main shock. Here, the difference in sound speeds in the ICM and non-thermal filaments plays an important role. Such conditions might be favorable for producing harder spectra than for a single weak shock. Since the ICM in this region can also be enriched with relativistic electrons, the problem of the particles acceleration from the thermal pool is alleviated too. 

\item One additional note on the data analysis: stray light (characteristic for Wolter~I X-ray mirrors) makes a significant contribution to the observed flux in the Coma outskirts. In the appendix, we provide an approximate model of the stray light in the 0.3-2.4~keV band, calibrated using observations of bright compact sources.
\end{itemize}

Overall, the Coma cluster represents a natural configuration of a minor merger, with the subcluster (the NGC~4839 group) located near the apocenter of its first orbit. The contact discontinuity separating the group ICM from the Coma gas and the leading shock are the natural outcomes of the merger. However, the morphology of the radio and X-ray images is more complicated than expected in the simplest version of this scenario. We suggest that magnetic-field-dominated filaments can explain some of the radio relic properties in the Coma cluster.

\section{Acknowledgments}
We are grateful to our referee for the helpful report. We thank Larry Rudnick for providing us with the 352-MHz FITS image of the Coma cluster published in \cite{2011MNRAS.412....2B} and Klaus Dolag for helpful discussions.

This work is partly based on observations with the eROSITA telescope on board \textit{SRG} observatory built by Roskosmos in the interests of the Russian Academy of Sciences represented by its Space Research Institute (IKI) in the framework of the Russian Federal Space Program, with the participation of the Deutsches Zentrum für Luft- und Raumfahrt (DLR). The eROSITA X-ray telescope was built by a consortium of German Institutes led by MPE, and supported by DLR. The \textit{SRG} spacecraft was designed, built, launched and is operated by the Lavochkin Association and its subcontractors. The science data are downlinked via the Deep Space Network Antennae in Bear Lakes, Ussurijsk, and Baikonur, funded by Roskosmos. The eROSITA data used in this work were converted to calibrated event lists using the eSASS software system developed by the German eROSITA Consortium and analysed using proprietary data reduction software developed by the Russian eROSITA Consortium.

Based on observations obtained with Planck \verb#(http://www.esa.int/Planck)#, an ESA science mission with instruments and contributions directly funded by ESA Member States, NASA, and Canada.

Some of the results in this paper have been derived using the HEALPix \citep{2005ApJ...622..759G} package.

Modeling of relativistic particles by A.M.B. at the Joint Supercomputer Center JSCC RAS and at the ``Tornado'' subsystem of the St.~Petersburg Polytechnic University supercomputing center was supported by the RSF grant 21-72-20020.
IK acknowledges support by the COMPLEX project from the European Research Council (ERC) under the European Union’s Horizon 2020 research and innovation program grant agreement ERC-2019-AdG 882679.

\bibliographystyle{aa} 
\bibliography{references}


\appendix
\section{Provisional stray light model}
\sloppypar
\label{app:stray}
Here, we describe a provisional stray light model that was used to make a first-order correction of the Coma cluster X-ray surface brightness profiles for the stray light contribution. This model extends the telescope point spread function (PSF) to large radii, beyond the central part, which is dominated by "nominal" events produced by two successive scatterings of incoming photons by the telescope mirror shells. The far outer wings of the PSF are formed primarily by singly scattered photons.  To simplify the problem, only the 0.4-2.3~keV band was considered and it was assumed that the all-sky survey data can be used to model the stray light during the scanning-mode observations of the Coma cluster.

To calibrate the entire PSF, including the core and the far wings, observations of different sources are needed. Indeed, very bright objects are needed to detect faint wings of the PSF.  However, for these objects, the core of the PSF will suffer from a heavy pile-up, dead-time, and various other adverse effects. On the other hand, for fainter sources, only the core of the PSF can be calibrated sufficiently well. To overcome this problem, a "ladder" of faint, medium, and bright sources has been used. The radial profiles, accumulated in the sky survey for groups of sources with different fluxes, were stitched together by matching their amplitudes over the range of radii, where adjacent groups provide robust measurements. 

\begin{figure}[h!]
\centering
\includegraphics[angle=0,trim= 0mm 4cm 0mm 1cm,width=1\columnwidth]{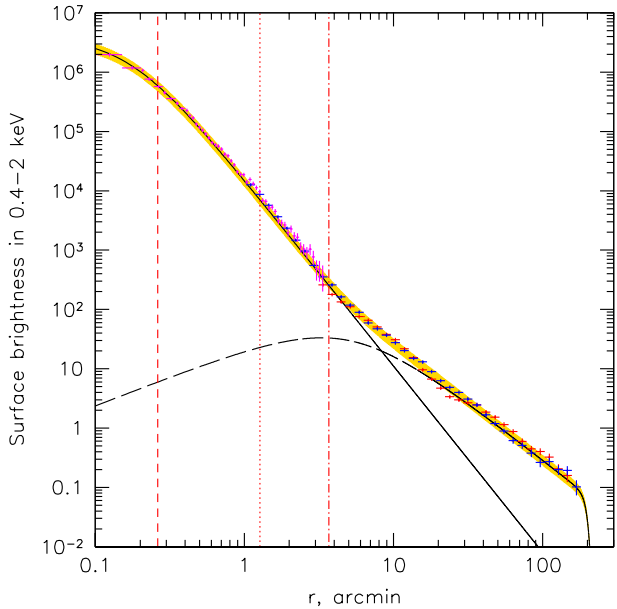}
\caption{Provisional model of the radial profile of the survey PSF that includes the contribution of the stray light (the yellow thick line). Crosses correspond to samples of faint (in the core) and bright (outer regions) compact sources used to estimate the PSF at different offsets. These samples were renormalized using overlapping radial ranges. The solid and dashed black lines correspond to the "core" and "extended" (= stray light) PSF components, respectively. This separation into two components is rather arbitrary, especially at radii $5-20'$. It was partly motivated by the desire to keep the simple functional form of the 'core' PSF over a broad range of radii \citepalias[as in][]{2021A&A...651A..41C}. Radii of circles encompassing  50\%, 90\%, and 97\%  of the 'core' PSF are marked with red lines.
}
\label{fig:psf_total}
\end{figure}
\FloatBarrier

Figure~\ref{fig:psf_total} shows the resulting azimuthly-averaged profile. The data corresponding to the different flux groups are shown with different colors. At the largest radii, the data points correspond to extremely bright sources, in particular, Cygnus X-1 and Cygnus X-2. The thick yellow line is the total survey PSF (${\rm P_t(r)}$). Two black curves correspond to the core PSF (${\rm P_c(r)}$) and the extended (stray-light) PSF (${\rm P_e(r)}$), respectively. The former is essentially an extrapolation (with minor modifications) of the central part of the PSF \citepalias[see Appendix B in][]{2021A&A...651A..41C} from $4'$ to large radii, while the dashed line is the analytic approximation of the extended part of the PSF needed to match the total PSF, namely,  ${\rm P_t(r)}\approx {\rm P_c(r)}+{\rm P_e(r)}$. We note, however, that this separation is rather arbitrary, especially at radii of $\sim 5-20'$. 

Analytic approximations for the shapes of the curves shown in Fig.~\ref{fig:psf_total} are as follows:
\begin{eqnarray}
P_c(r) &=& \frac{1}{~~~\left ( 1+\left [ \frac{r}{r_{c,c}}\right ]^2 \right)^{(3\beta_c-0.5) }},\\
\label{eq:psfc}
P_e(r) &=&\frac{3\times 10^{-5}}{~~~\left ( 1+\left [ \frac{r}{r_{c,e}}\right ]^2 \right)^{(3\beta_e-0.5) }}\times \frac{r}{r+r_{c,e}}\times e^{-\left (\frac{r}{r_{i,o}} \right )^{20}},
\label{eq:psfe}
\end{eqnarray}
where $r_{c,c}=0.17'$, $\beta_c=0.69$, $r_{c,e}=5'$, $\beta_e=0.5$, $r_{i,o}=200'$. The last two terms in Eq.~\ref{eq:psfe} are used to suppress the contribution of the extended PSF at small radii and introduce a sharp cutoff at $r\approx 3^\circ$. Only the relative normalization of these components matters. In particular, these two components may not reflect the contributions of the single and double-scattered photons at any given radius. However, at small (large) radii, double (single) scatterings are dominant.

We reiterate here that the above expressions are approximate and might change in the future. For the functional forms used in Eq.~\ref{eq:psfc} and \ref{eq:psfe}, the total flux in the extended component is $\sim6\%$ of the compact PSF. This means that, on average, 6\% of the photons at any given position in the survey maps are coming from regions rather far (tens of arcminutes - up to three degrees) from this position. 

\begin{figure}
\centering
\includegraphics[angle=0,trim= 0mm 0cm 0mm 0cm,width=1\columnwidth]{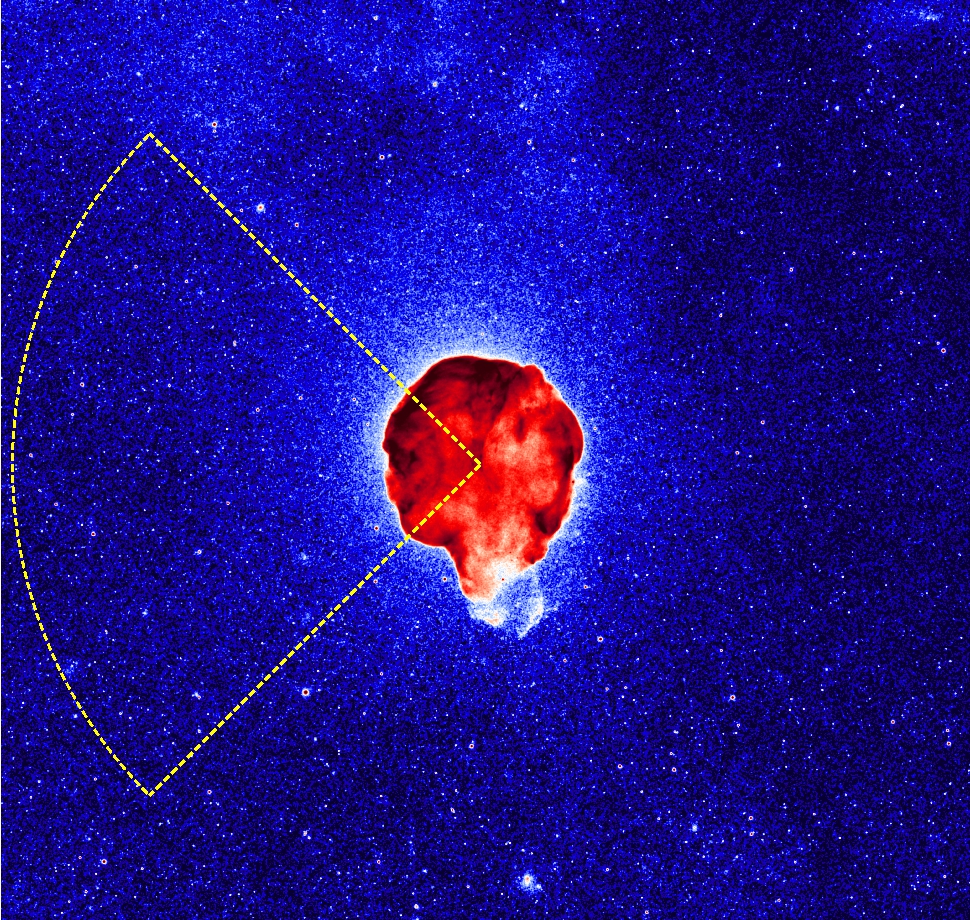}
\caption{Visualization of stray light around bright supernova remnant Cygnus Loop. A white glow around the remnant is largely due to the stray light.   
}
\label{fig:cygloop_image}
\end{figure}

\begin{figure}
\centering
\includegraphics[angle=0,trim= 0mm 4cm 0mm 1cm,width=1\columnwidth]{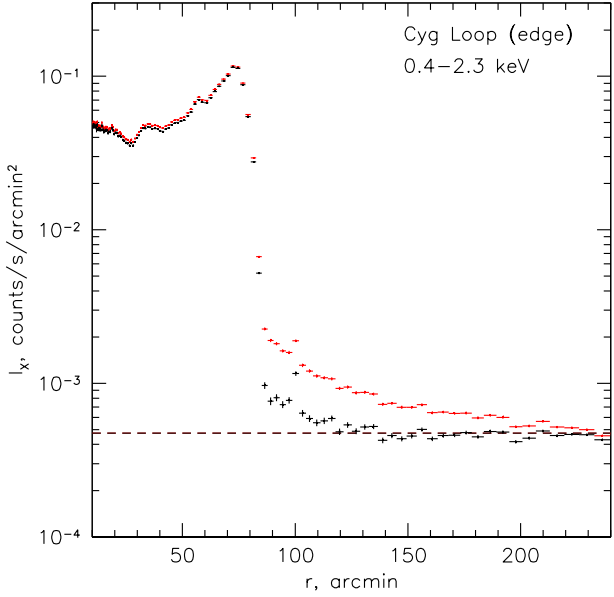}
\caption{Illustration of the first-order stray-light correction for the Cygnus Loop. Red points correspond to the radial profile in the wedge shown in Fig.~\ref{fig:cygloop_image}. Black points show the same profile after 1-st order correction for the stray light. The dashed horizontal line is the flux level away from the Cygnus Loop.  
}
\label{fig:cygloop_radial}
\end{figure}

The first-order correction of the stray-light contribution is possible with the following simple recipe. The observed surface brightness can be represented as a convolution of the true surface brightness $I_X$ with the total PSF:
\begin{eqnarray}
I_{obs}=I_X\ast P=I_X\ast P_c+I_X\ast P_e,
\end{eqnarray}
where $\ast$ denotes a convolution. The contribution of the stray light can be estimated by convolving the observed image with the extended PSF:
\begin{eqnarray}
I_{stray}\approx (I_{obs}-I_{mean})*P_e=I_X\ast P_c \ast P_e+ I_X\ast P_e \ast P_e \approx \nonumber \\ I_X\ast P_c \ast P_e \approx I_X\ast P_e,
\label{eq:istray}
\end{eqnarray}
where the $I_X\ast P_e \ast P_e$ is neglected since its contribution is second-order in terms of the small parameter $\int P_e/\int P_t$. We note that the mean sky level $I_{mean}$ has been subtracted.  
As a final step, a corrected image is calculated as: 
\begin{eqnarray}
I_{corr}\approx I_{obs}-I_{stray}.
\end{eqnarray}

This approach was further validated using the very bright galactic supernova remnant (SNR) Cygnus Loop, which has very sharp boundaries. Fig.~\ref{fig:cygloop_image} shows the SNR image obtained in the all-sky survey. A white "glow" around the boundaries of the images is due to the stray light. The corresponding radial profile is shown in Fig.~\ref{fig:cygloop_radial}. For comparison, the black points show the same profile for the corrected image. On large scales, the correction works well, effectively suppressing the contribution of the stray light. Close to the sharp boundaries of the SNR, the residual signal is still visible. This is the result of our particular choice of the decomposition of the PSF into two components and second order effects not captured by Eq.~\ref{eq:istray}. 

\begin{figure}
\centering
\includegraphics[angle=0,trim= 0mm 0cm 0mm 0cm,width=1\columnwidth]{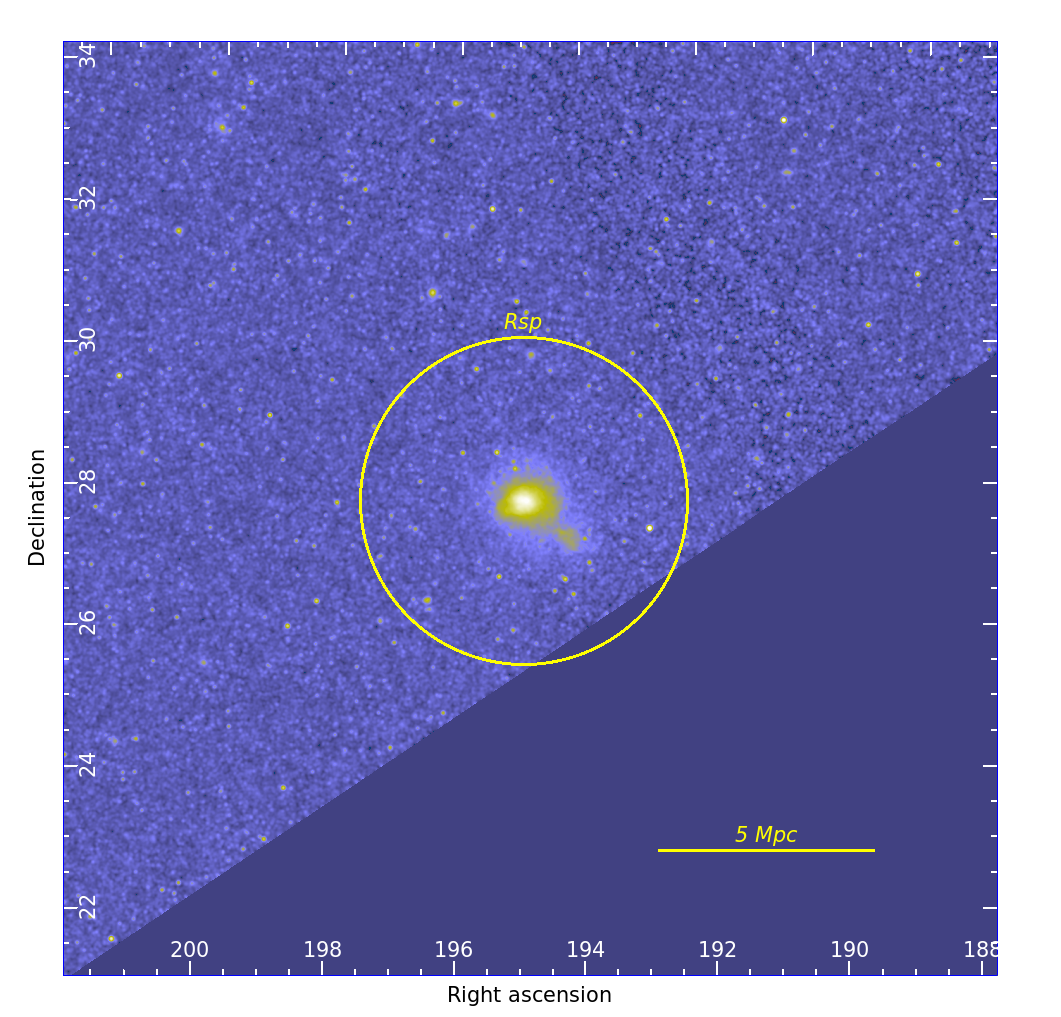}
\caption{$13.3\times 13.3$ deg field centered at the Coma cluster in the 0.4-2.3 keV band based on the all-sky survey data. This image was smoothed with the $\sigma=60"$ Gaussian. The yellow circle has a radius of 140 arcmins. 
}
\label{fig:coma_sur}
\end{figure}

\begin{figure}
\centering
\includegraphics[angle=0,trim= 0mm 0cm 0mm 0cm,width=1\columnwidth]{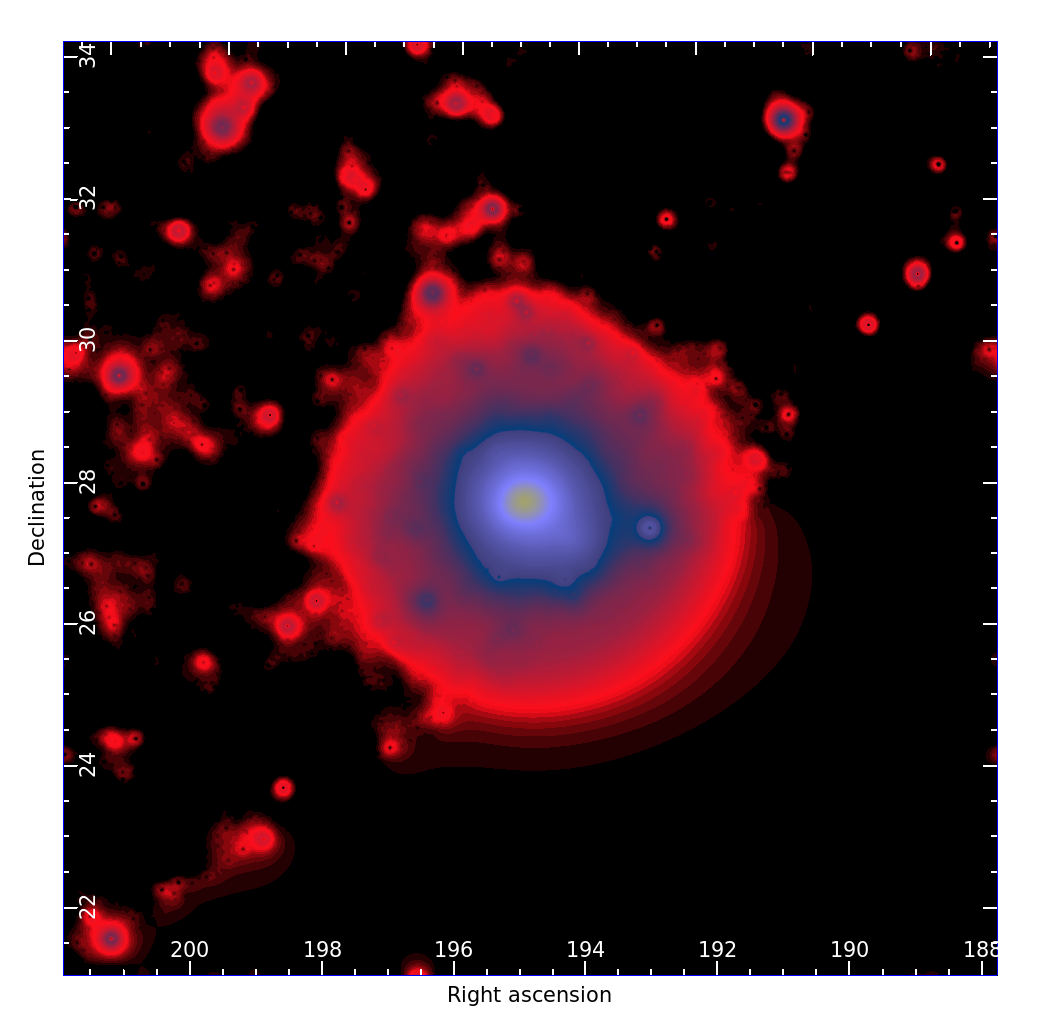}
\caption{Coma field image shown in Fig.~\ref{fig:coma_sur} convolved with $P_e(r)$ according to Eq.~\ref{eq:istray}.
The scale has been changed by a factor of 10 (upper boundary) compared to Fig.~\ref{fig:coma_sur}. 
}
\label{fig:coma_stray}
\end{figure}

The 0.4-2.3~keV image of the Coma field (Fig.~\ref{fig:coma_sur})  obtained in the course of the all-sky survey  has been used to evaluate $I_{stray}$, as shown in  Fig.~\ref{fig:coma_stray}. This image is used in this study as an additional background component, which is subtracted from the raw image in order to get radial profiles corrected for the contribution of stray light.

\label{lastpage}
\listofobjects

\end{document}